**Type of manuscript**

Research Article

**Manuscript title**

Why recommended visit intervals should be extracted when conducting longitudinal analyses using electronic health record data: examining visit mechanism and sensitivity to assessment not at random

**Short title (up to 70 characters):**

Using recommended visit intervals in longitudinal analyses of EHR data

**Authors and affiliations**

Rose H. Garrett[1,2], Masum Patel[1], Brian M. Feldman[1,3,4], Eleanor M. Pullenayegum*[1,2]

[1]Child Health Evaluative Sciences, The Hospital for Sick Children, Toronto, Ontario, Canada

[2]Division of Biostatistics, Dalla Lana School of Public Health, University of Toronto, Toronto, Ontario, Canada

[3]Division of Rheumatology, Department of Pediatrics, Faculty of Medicine, University of Toronto, Toronto, Ontario, Canada

[4]Institute of Health Policy, Management and Evaluation, Dalla Lana School of Public Health, University of Toronto, Toronto, Ontario, Canada

## Abstract

Electronic health records (EHRs) provide an efficient approach to generating rich longitudinal datasets. However, since patients visit as needed, the assessment times are typically irregular and may be related to the patient's health. Failing to account for this informative assessment process could result in biased estimates of the disease course. In this paper, we show how estimation of the disease trajectory can be enhanced by leveraging an underutilized piece of information that is often in the patient's EHR: physician-recommended intervals between visits. Specifically, we demonstrate how recommended intervals can be used in characterizing the assessment process, and in investigating the sensitivity of the results to assessment not at random (ANAR). We illustrate our proposed approach in a clinic-based cohort study of juvenile dermatomyositis (JDM). In this study, we found that the recommended intervals explained 78% of the variability in the assessment times. Under a specific case of ANAR where we assumed that a worsening in disease led to patients visiting earlier than recommended, the estimated population average disease activity trajectory was shifted downward relative to the trajectory assuming assessment at random. These results demonstrate the crucial role recommended intervals play in improving the rigour of the analysis by allowing us to assess both the plausibility of the AAR assumption and the sensitivity of the results to departures from this assumption. Thus, we advise that studies using irregular longitudinal data should extract recommended visit intervals and follow our procedure for incorporating them into analyses.

# 1 INTRODUCTION

Electronic health records (EHRs) are being increasingly used as a low-cost approach to investigate the progression of diseases and the impact of interventions over time. However, since there are no study-specific visits, and patients interact with the healthcare system on an as-needed basis, the frequency and timings of assessments are highly variable and are often related to the patient's health status. The intervals between visits vary both between and within patients, and the physician typically recommends that sicker patients visit the clinic more frequently. Thus, the assessments are not only irregularly spaced, but are also more frequent when patients are sicker.

Over-representation of assessments when the patient is unwell results in distorted estimates of the disease trajectory over time.[1] For example, in a study of HIV-positive new mothers, Bůzková et al.[2] found that ignoring the overrepresentation of measurements on sicker individuals resulted in two-fold overestimation of pneumonia prevalence. Thus, it is crucial to account for the irregular visit times.

The first step in accounting for visit irregularity is to characterize the assessment mechanism. In this paper we shall assume that the outcome is assessed at every visit, and thus the terms "visit" and "assessment" are used interchangeably; more generally the outcome may be assessed at only a subset of visits. The classic missing data mechanism taxonomy proposed by Rubin[3] can be adapted to the irregular assessment context as follows (see [1,4]): scenarios where the assessment and outcome processes are independent are defined as assessment completely at random (ACAR); Assessment at random (AAR) occurs when visiting at time t is independent of the outcome at time t after conditioning on the observed data before time t (which includes the recorded measurements of the outcome, the assessment times, and other covariates); assessment not at random (ANAR) refers to scenarios where visiting at time t is not independent of the outcome at time t, even given all the recorded data before time t. While statistical methods exist for handling AAR, [2, 5-7] there are no methods that can handle a general case of ANAR.

It is therefore essential to consider how plausible the AAR assumption is, and the likely impact of ANAR. In the context of missing data, global sensitivity analysis methods have been developed to investigate the impact of missing not at random on the results over a range of possible assumptions about how the value of the outcome affects missingness.[8-15] Global sensitivity analysis for irregular longitudinal data is currently under-explored, but would similarly involve assessing the possible impact of ANAR over a range of plausible assumptions about the relationship between the assessment process and the outcome. Although under-used in practice, [16] sensitivity analysis has been identified as a crucial analytic component by several reporting guidelines.[17-22]

Smith et al.[23,24] recently developed methodology for conducting sensitivity analysis of irregularly observed longitudinal data in the context of clinical trials where there were pre-specified visit times. We propose to extend their approach to the EHR data setting, where there are no common visit times, by incorporating the concept of visit categories, and developing a framework for identifying a visit as in-window, early or late.

We propose a simple solution to assessing plausibility of AAR and sensitivity to ANAR in EHR data: to extract the physician's recommendation on when the next visit should occur. This information is often in the EHR but is typically not included when creating datasets and consequently has hitherto not been used for analysis. The recommended intervals form a patient-specific visit schedule (which updates at each visit), giving a benchmark for when visits should plausibly occur, similar to the pre-specified visit times in clinical trials. If patients adhere perfectly to the recommended visit interval, then after accounting for the recommended interval, no other factors are associated with the visit times, and therefore AAR will hold: this is not ACAR because the visit interval is determined based on patient factors, nor is it ANAR because the visit interval suggested is done at the prior visit and therefore cannot be influenced by the outcome at the subsequent visit.

The greater the deviations between actual and recommended intervals, the more scope there is for bias due to ANAR. Moreover, knowing the recommended visit intervals allows visits to be classified as early, late, or within a reasonable window, making the extension of global sensitivity analysis to EHR data applications possible.

In this paper we demonstrate how the physician-recommended intervals between visits can be used both to characterize the assessment process and to examine the sensitivity of results to specific violations of AAR. Specifically, we 1) illustrate how observed and recommended visit intervals can be compared to assess the extent of irregularity potentially due to unobserved patient factors, and 2) develop a framework for using recommended visit intervals to investigate the potential impact of specific cases of ANAR on inferences via global sensitivity analysis.

We begin this paper by describing our motivating example. Next, in Section 3 we discuss recommended visit intervals in more depth and show how to use them to assess the plausibility of AAR. Then in Section 4, we show how to fit an initial model for the disease trajectory that assumes AAR. Lastly, in Section 5 we show how recommended visit intervals can be used for global sensitivity analysis when ANAR is suspected.

## 2 MOTIVATING EXAMPLE: THE JDM STUDY

We consider a previously reported clinic-based cohort study of patients with juvenile dermatomyositis (JDM).[25] JDM is a rare, often chronic, autoimmune disease; common symptoms include rashes and muscle weakness. JDM manifests itself differently across individuals, and the severity of symptoms also varies within individuals over time as medications take effect; thus, from the outset, we expect to see irregularly spaced observations in this dataset. In addition, flares are a feature of the disease course and require prompt treatment, which consequently leads to patients visiting the clinic earlier than expected. This patient-driven visit pattern leads to concern over ANAR.

This study was approved by The Hospital for Sick Children (SickKids) Research Ethics Board (REB #1000019708). SickKids is a paediatric hospital in Toronto, Ontario, Canada. The study population consisted of patients enrolled in the specialized JDM clinic at SickKids who visited the clinic at least twice between June 1, 2000 and May 31, 2018, and had a documented date of diagnosis in their chart (note that this cohort included patients who had dates of diagnosis and some clinic visits preceding June 1, 2000). Data were collected from patient charts within the EHR database.

Our sample included 149 patients with a combined total of 2912 visits (see Supplemental Figure S1 for a flow chart describing the sample size and final number of observations in the dataset, and Supplemental Tables 1 and 2 for patient and visit-level characteristics of the sample). Patients were enrolled in the cohort at diagnosis and followed as part of usual care. Of the 149 patients enrolled in the clinic after diagnosis, 84 (56%) were followed up until transitioning to adult care. The remaining patients either had their follow-up censored at the end of the study period, or transferred to another hospital, or follow-up was stopped for some other reason before the patient reached age 18.

The modified Disease Activity Score ($DAS_{mod}$) can be used to measure disease activity in patients with JDM.[26] The $DAS_{mod}$ ranges from 0 to 12, and a higher score indicates more severe disease activity. In our study, the target of inference is the progression of $DAS_{mod}$ over time, where time is measured in years since diagnosis. The standard of care was that $DAS_{mod}$ was assessed at every visit; thus, in this study, assessment time and visit time are equivalent.

Figure 1 depicts visit times from a randomly selected subset of the patients. From this plot, it is immediately clear that the visit times are highly irregular. This irregularity can lead to analytic challenges because the disease outcome model must account for the fact that the visit times are often related to the health status of the patient, and to further complicate matters, due to the occurrence of flares, ANAR may be present. Thus, in order to accurately estimate the disease trajectory over time, it is important to identify the sources of the irregularity, and also conduct global sensitivity analysis to investigate the potential impact of irregularity that cannot be explained by the observed data history.

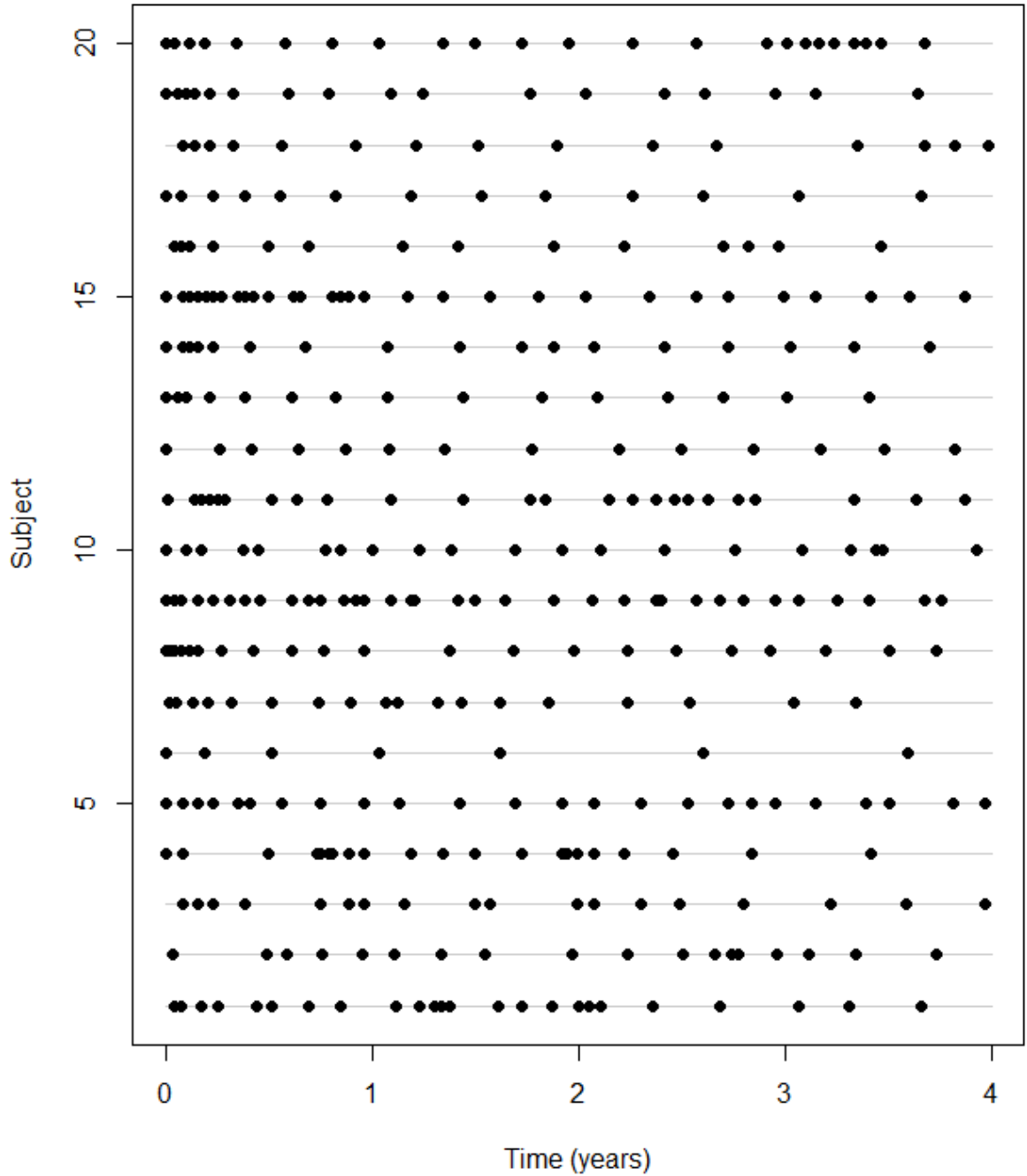

Figure 1: Visit times for a random subset of 20 patients in the JDM dataset over their first 4 years of follow-up since being diagnosed. Each dot represents a visit, and the grey solid lines indicate times where the patient was still being followed at the clinic, but there was no visit.

# 3 RECOMMENDED VISIT INTERVALS & AGREEMENT WITH OBSERVED INTERVALS

## 3.1 The role of recommended visit intervals & their extraction from EHRs

The irregularity in assessment times in observational cohort studies arises from two sources: the variability in physician recommendations on when patients should next visit the clinic, and the variability in how closely patients adhere to these recommendations. It is important to distinguish between physician-driven and patient-driven irregularity because they have different analytic implications. More specifically, variability in recommended visit intervals leads to AAR, while discrepancies between observed and recommended intervals are likely due to unobserved factors and hence opens up the possibility of ANAR.

Therefore, the recommended visit intervals are key in both disentangling the patient-driven irregularity in assessment times from the physician-driven irregularity, and as a means to operationalizing the global sensitivity analysis.

Despite their usefulness, recommended visit intervals are unlikely to be available as a predefined variable with measurements that have already been neatly organized and collected, since the physician recommendations are not typically used in analysis. Nevertheless, we conjecture that the recommended visit intervals should generally neither be difficult nor expensive to extract. For example, in the JDM study, this information was often easily discernible from documentation in the patient's EHR.

In more detail, in the JDM study, letters to the primary care physician and physician notes were used to extract the recommended intervals between visits, and comments regarding follow-up times were also extracted to provide additional information, such as explanations as to why some recommended follow-up times were missing.

The specialized JDM clinic at SickKids generally runs every two weeks, so the standard recommended intervals are multiples of two weeks and the typical shortest possible recommended interval would be two weeks, however patients can come in sooner if more urgent care is needed (e.g., in the case of an adverse event such as a disease flare, medication complication, or unexpected symptoms). There is no universally recommended follow-up interval; instead, the recommended follow-up time depends on individual patient needs that change over time (e.g., disease severity, medication monitoring, heterogeneity in different patient disease trajectories over time). For the 14.6% (426) of the 2912 total visits in the JDM dataset that were missing recommended intervals, the reason for missingness was extracted whenever possible. For example, the next visit may be "pending family decision", "pending MRI", or "pending consultation with another doctor." We also note that 13.5% (392) of the visits were missing $DAS_{mod}$, and 2.6% (77) were missing both the recommended interval and $DAS_{mod}$.

## 3.2 Results: assessing agreement between observed and recommended intervals

Figure 2 shows the difference between the observed and recommended interval versus the recommended interval. The area of the plot enclosed by the 25th and 75th quantiles shows where the central 50% of the data lie. For example, for a recommended interval of 1 month, the visit occurred between 2.4 days earlier and 11.4 days later than recommended in 50% of the observations. Similarly, the area of the plot enclosed by the 5th and 95th quantiles shows where the central 90% of the data lie. In 90% of cases for which the recommended interval was 1 month, the visit occurred between 11.4 days earlier and 1.7 months later than recommended.

Figure 3 shows the ratio of observed to recommended interval versus the recommended interval. For illustration, consider a recommended interval of 12 months. In 50% of the observations, the ratio lay between 0.98 and 1.1, and for 90%, the ratio fell between 0.49 and 1.3.

Using the median absolute deviation (MAD) as a measure of variability, we found that the recommended visit intervals explained an estimated 78% of the variability in the observed intervals (adjusted MAD=0.30 months, unadjusted MAD =1.38 months)[i].

In general, close agreement between the observed and recommended intervals raises the possibility that ANAR has a small effect on estimated trajectories. Thus, we conduct analysis under AAR as an initial step before undertaking global sensitivity analysis. More specifically, we use IIW-GEEs, which assume AAR, to model the disease trajectory over time.[5] To fit an IIW-GEE, the usual GEE is weighted by the inverse of the visit intensity, which can be thought of as the instantaneous incidence of visiting at any given time, conditioning on the observed history up to this time.

---

[i] We quantified the amount of variability in observed intervals between visits explained by the recommended intervals by comparing the median absolute deviation (MAD) in the observed intervals to the MAD, accounting for the recommended intervals. We fit a median regression model with the observed interval between visits as the response and recommended interval as the predictor. To calculate the adjusted MAD, we computed the median of the absolute deviations of the observed intervals from their fitted medians.

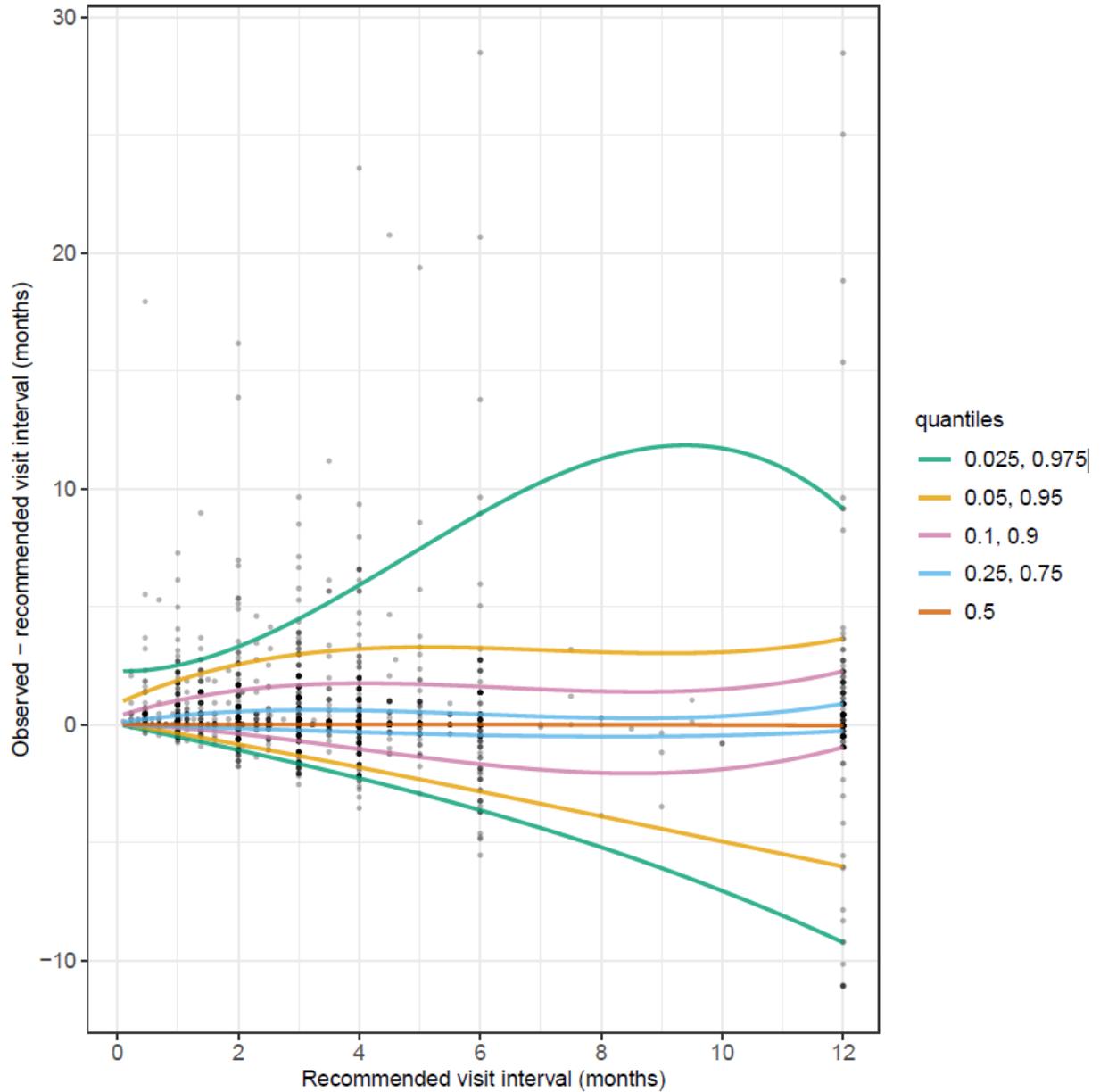

Figure 2: Difference between observed and recommended visit interval (in months) vs. the recommended visit interval (in months). The trend lines represent quantile regressions on the difference between observed and recommended visit interval, with a spline function of the recommended visit interval as the predictor. Note that the single visit with a recommended interval of 14 months was excluded from the plot for clarity of presentation. We used the ggplot2 R package[27] to fit the quantile regressions and produce the plot.

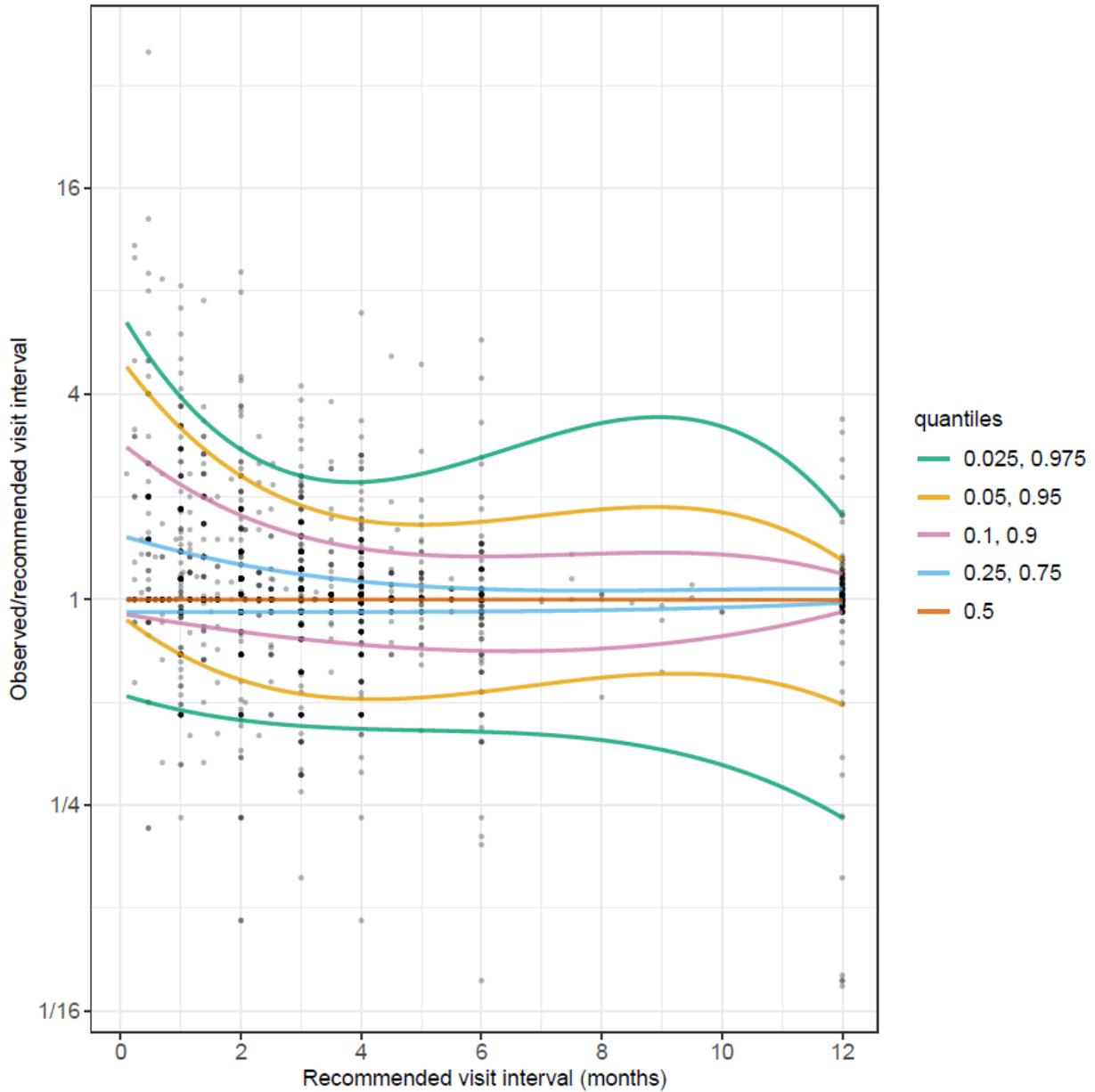

Figure 3: Ratio of observed to recommended interval vs. recommended interval (in months), with y-axis shown on the log-scale. The trend lines represent quantile regressions on ratio of observed to recommended interval, with a spline function of the recommended interval as the predictor. Note that the single visit with a recommended visit interval of 14 months was excluded from the plot for clarity of presentation. We used the ggplot2 R package[27] to fit the quantile regressions and produce the plot.

# 4 ANALYSIS UNDER AAR

In this section, we briefly review the IIW-GEE as we will build upon this modelling technique in Section 5 when we consider global sensitivity analysis.

## 4.1 Notation

We can define the visit intensity for individual i at time since diagnosis t in a general form as follows:

$$\lambda_i(t; F_i(t)) = \lim_{\delta \to 0} \frac{P(N_i(t+\delta) - N_i(t) = 1 \mid F_i(t))}{\delta}$$

(1)

where $F_i(t)$ is some arbitrary set of variables and $N_i(t)$ is the counting process for the visit times for subject $i$.

We note that in this paper, we use the following notation for any arbitrary process $A$: $\Delta A(t) = A(t) - A(t^-)$ with $t^-$ being the instant of time right before $t$. That is, $A(t^-) = \lim_{s \uparrow t} A(s)$. We let $T_{ij}$ denote the time of the j[th] visit for subject i, and we can also use the simplified notation $A_{ij} = A_i(T_{ij})$.

We let $Y_i(t)$ denote the outcome for individual i at time t, and we set $E(Y_i(t)|X_i(t)) = X_i(t)\beta$ as the marginal model for the disease outcome (the model of clinical interest), where $X_i(t)$ is a vector of possibly time-varying covariates.

We let $\mathcal{H}_i(t)$ denote the observed history for individual *i* up to time *t,* which would typically include the recorded measurements of the outcome, the visit times, and other (possibly time-varying) covariates.

The most commonly used visit intensity model is as follows:

$$\lambda_i(t; \mathcal{H}_i(t)) = \lambda_0(t)\exp(Z_i(t)\gamma)$$

(2)

where $\lambda_0(t)$ is a baseline intensity function and $Z_i(t) \subset \mathcal{H}_i(t)$. For example, in the JDM data example, we take $Z_i(t)$ to be the recommended interval to the next visit assigned to individual *i* at the last visit before time t.

We let $\tau$ denote the end of the study period.

## 4.2 Modelling equations

The IIW-GEE equations, using an independent working-correlation structure, are:[5,28]

$$\sum_i \int_0^\tau w_i(t) X_i(t)' (Y_i(t) - X_i(t)\beta) dN_i(t) = 0$$

(3)

with the inverse-intensity weight $w_i(t) = 1/\lambda_i(t)$.

Models under AAR take $F_i(t) = \mathcal{H}_i(t)$ as the conditioning set in Equation 1.

## 3.4 Modelling the JDM dataset: IIW-GEE assuming AAR

In this section, we fit an IIW-GEE to the JDM dataset. This approach, which assumes AAR, serves as a starting point for drawing inferences about the disease trajectory over time. Throughout this section, we briefly discuss modelling choices and results that will be explained in more depth in the global sensitivity analysis portion of this paper, where we will look at how inferences change under departures from this benchmark AAR assumption.

For the analysis of the JDM dataset, we also assume that given the time of the last visit and the recommended interval assigned at the last visit, the visit intensity is independent of other components of the observed history. We justify this assumption as being reasonable given our previous findings of close agreement between the observed and recommended visit intervals.

In more detail, we modelled the visit intensity using a piecewise exponential survival model, allowing the log intensity to depend on a smoothing spline function $S$ of the recommended visit interval. The model was piecewise in that a separate baseline hazard was estimated based on the type of visit. That is, we partitioned the visit time into intervals of time at risk of five different types of visits: early, very early, late, very late, and in-window. This visit classification system was formulated using clinical expertise, with very early visits being defined as occurring more than one month earlier than recommended by the physician, early visits between two weeks and one month earlier than recommended, very late visits more than two times later than recommended, late visits between 1.5 and two times later than recommended, and the in-window group consisted of visits that were neither early/very early nor late/very late (see Figure 4).

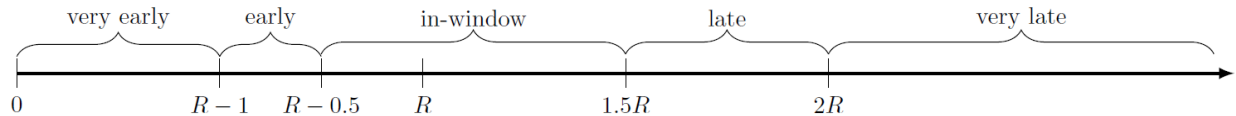

Figure 4: Visit categories, shown on the month scale (i.e., 2 weeks is represented by half a month). R represents the recommended visit interval. Note that this diagram is not drawn to scale, but is based on an example recommended interval of 2 months.

We note that of the 2345 visits in the JDM dataset with both a recorded recommended visit interval and uncensored observed interval, 77.1% (1809) were classified as in-window, 6% (140) very early, 5.8% (137) very late, 5.8% (135) late, and 5.3% (124) early.

Mathematically, the visit intensity model is as follows:

$$\lambda_i(t; R_{iN_i(t^-)}, T_{iN_i(t^-)}) = \lambda_{V_i(t)} \exp\{S(R_{iN_i(t^-)})\gamma\}$$

(4)

where $R_{ij}$ is the recommended interval between the $j^{th}$ and $j+1^{st}$ visit for individual $i$, assigned by the physician at the end of visit $j$, and the notation $R_{iN_i(t^-)}$ then denotes the recommended interval assigned to individual $i$ at the last visit before time $t$. $V_i(t)$ denotes the visit category for the potential visit occurring at time $t$ for subject $i$. More formally, using the time scale of months, we have $V_i(t) = $ very early if $t - T_{iN_i(t^-)} < R_{iN_i(t^-)} - 1$, $V_i(t) = $ early if $R_{iN_i(t^-)} - 1 \leq t - T_{iN_i(t^-)} < R_{iN_i(t^-)} - 0.5$, $V_i(t) = $ late if $1.5R_{iN_i(t^-)} < t - T_{iN_i(t^-)} \leq 2R_{iN_i(t^-)}$, $V_i(t) = $ very late if $t - T_{iN_i(t^-)} > 2R_{iN_i(t^-)}$, and $V_i(t) = $ in-window otherwise. We note that it is not possible to have early or very early visits for some of the smaller-sized recommend visit intervals.

To obtain the weights for the IIW-GEE, we invert the visit intensities.

For the marginal outcome model, $E(DAS_{mod_i}(t)) = \boldsymbol{X}_i(t)\boldsymbol{\beta}$, we took $\boldsymbol{X}_i(t)$ to be a cubic smoothing spline basis for the time since diagnosis, noting that we do not have subject-specific covariates for this analysis.

For reference we also included an unweighted GEE, and both GEEs used working independence and a linear link function. For each model, we calculated the area under the curve (AUC) when the estimated mean disease activity was plotted against time since diagnosis.

We assume unknown observations are missing at random. We imputed the missing values using a Bayesian model, drawing 30 sets of imputations. We report on analysis using complete cases in Supplemental Figures S2 and S3.

All analyses in this paper were conducted using R version 4.1.0.[29] See Appendix D in the Supporting Information for further modelling details and code.

Figure 5 shows that the disease activity trajectory estimated by the unweighted GEE (black curve) is generally positioned higher on the y-axis compared to the IIW-GEE under AAR (red curve).

Moreover, although there was close agreement between the observed and recommended visit intervals, which supports the plausibility of AAR, there was still some unexplained variability in visit timings that could potentially be patient-driven, and there are good clinical reasons to believe this is related to disease activity. Thus, it is pertinent to investigate possible cases of patient-driven ANAR and their impact on the estimated disease trajectory curve.

In the next section, we begin the global sensitivity analysis portion of the paper, where we first review existing methodology from the literature, and then introduce our novel approach and its application to the JDM dataset.

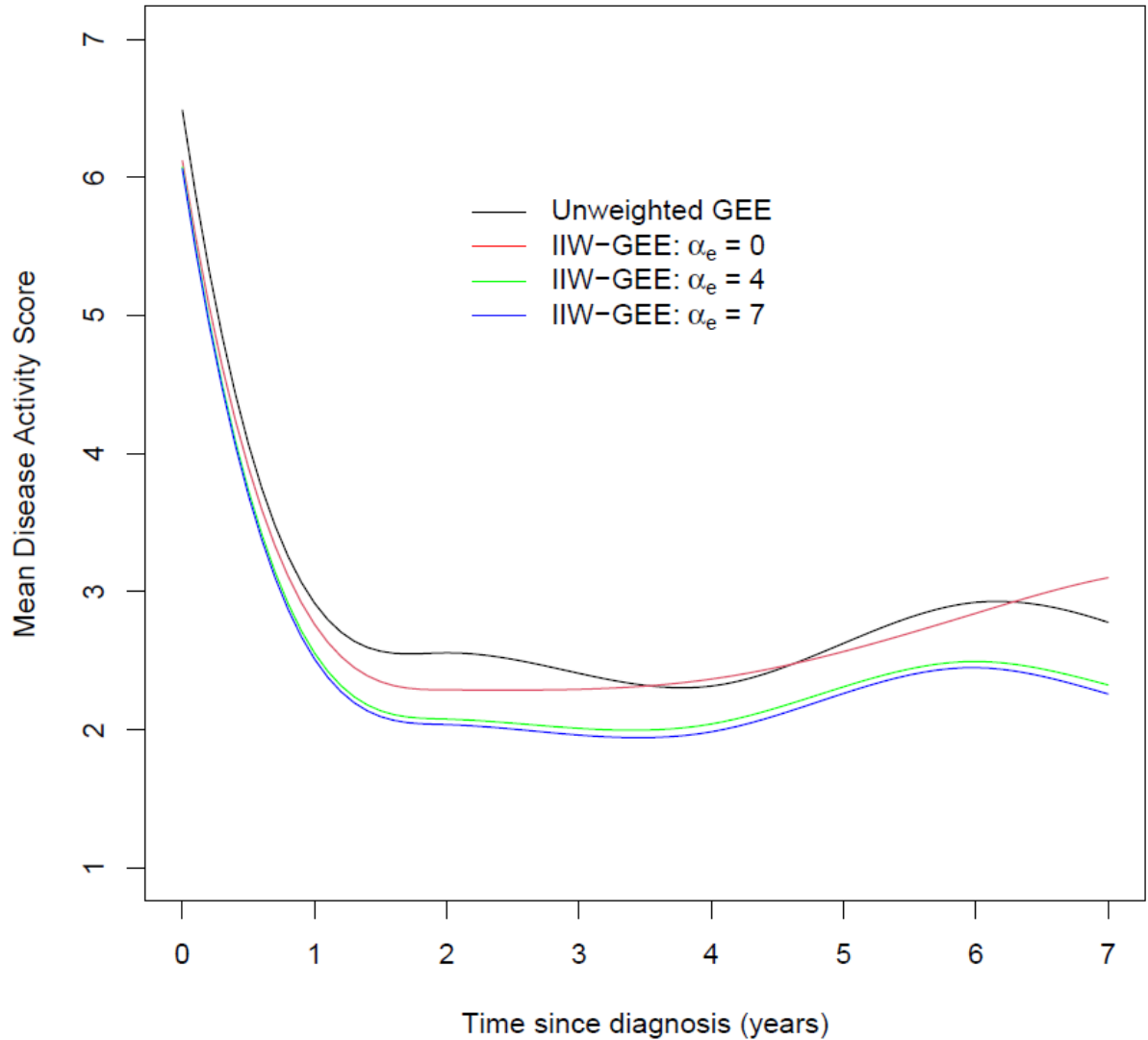

Figure 5: Fitted expected disease activity as a function of time using unweighted and inverse-intensity weighted (IIW) GEEs. The association parameter quantifying the strength of the association between the hazard of visiting earlier than recommended and worsening health condition, $\alpha_e$, is shown at 3 example values specified through the sensitivity analysis. The $\alpha_l$ parameter representing the strength of the association between the hazard of visiting later than recommended and worsening condition is set to be 0. The red curve (with $\alpha_e = \alpha_l = 0$) represents the IIW-GEE assuming AAR.

# 5 USING RECOMMENDED VISIT INTERVALS FOR GLOBAL SENSITIVITY ANALYSIS WHEN ANAR IS SUSPECTED

## 5.1 Global sensitivity analysis in clinical trials with irregular observation

The sensitivity of the disease trajectory estimated through the IIW-GEE to varying degrees of violations of the AAR assumption can be assessed through global sensitivity analysis. To formulate the visit intensity model under the specified cases of ANAR, exponential tilting can be used. Smith et al. [23,24] formulate the exponential tilting assumption in the context of irregular observation in clinical trials as follows:

$$f\big(Y_i(t) \mid \Delta N_i(t) = 0, \mathcal{H}_i(t)\big) = f\big(Y_i(t) \mid \Delta N_i(t) = 1, \mathcal{H}_i(t)\big) \frac{\exp\{q(Y_i(t), \mathcal{H}_i(t); \alpha)\}}{c(t, \mathcal{H}_i(t); \alpha)}$$

(2)

where $q(Y(t), \mathcal{H}(t); \alpha)$ is a user-specified "tilting" function which is equal to zero at $\alpha = 0$, with $\alpha$ representing a vector of parameters to be varied over a grid of user-specified plausible values in the sensitivity analysis, and $c(t, \mathcal{H}(t); \alpha) = E(\exp\{q(Y(t), \mathcal{H}(t); \alpha)\} \mid \mathcal{H}(t), \Delta N(t) = 1)$ is a normalizing constant.

Using Bayes' theorem, Smith et al.[23,24] show that the exponential tilting assumption yields the following visit intensity model allowing for ANAR:

$$\lambda_i\big(t; \mathcal{H}_i(t), Y_i(t)\big) = \lambda_i\big(t; \mathcal{H}_i(t)\big) \frac{c(t, \mathcal{H}_i(t); \alpha)}{\exp\{q(Y_i(t), \mathcal{H}_i(t); \alpha)\}}$$

(3)

where they also assume that the intensity function does not depend on unobserved variables other than the current outcome $Y_i(t)$ or on any future outcome variables occurring after time t.

Equation 5 provides the inverse-intensity weights to be plugged into the IIW-GEE. The γ parameter(s) introduced in Equation 2's formula for $\lambda_i\big(t; \mathcal{H}_i(t)\big)$ is estimable from the observed data, while α is varied over a user-specified range.

This approach links the distribution of outcomes for individuals who were not assessed at time t to the distribution for individuals with the same observed past who were assessed at time t through a tilting factor. The tilting factor alters the shape of the distribution by allocating more/less probability mass to smaller/larger values of $Y_i(t)$ relative to the distribution among individuals who were assessed at time t. The direction of the shift depends on the specific parametric form of the tilting function and values of $\alpha$ supplied. We note that this framework distinguishes between visiting versus not visiting at exactly time *t*, but for EHR data we may wish to apply less strict bounds around *t* by allowing patients who were assessed shortly prior to or after time *t* to have the same outcome distribution as those assessed exactly at time *t*, controlling for the observed history.

## 5.2 Using recommended visit intervals to conduct global sensitivity analysis

To see why the distinction between visiting versus not visiting at exactly time *t* can be too sharp in the context of EHR data, consider the following example: suppose there is a group of patients with a certain observed past who are advised to come back to the clinic in 6 months, and Subgroup 1 comes back in exactly 6 months, while Subgroup 2 comes back two weeks before the 6 month mark, and Subgroup 3 comes back two weeks after the 6 month mark. We would not expect there to be a difference in the distribution of outcomes for these three subgroups, since the slight variations in assessment timing are likely due to scheduling difficulties in busy clinics, other life priorities, and/or random chance, rather than true differences in health status. However, if we instead suppose there is a group of patients with a certain observed past who are advised to come back to the clinic in 12 months, we would expect there to be a difference in 6-month outcomes for patients who come back in 12 months' time (on-time subgroup) versus patients who come back in 6 months' time (early subgroup), since patients who visit 6 months earlier than originally recommended are likely doing so because they developed unexpected symptoms between scheduled appointments and require prompt treatment. The patients in the early subgroup would be assessed when they visited at 6 months while the patients in the on-time subgroup would not have been assessed yet at the 6 month mark, and we could hypothesize that the on-time subgroup's distribution would be tilted with more weight on lower values of Y(6 months) relative to the early subgroup. See Figure 6 for a graphical illustration of this example.

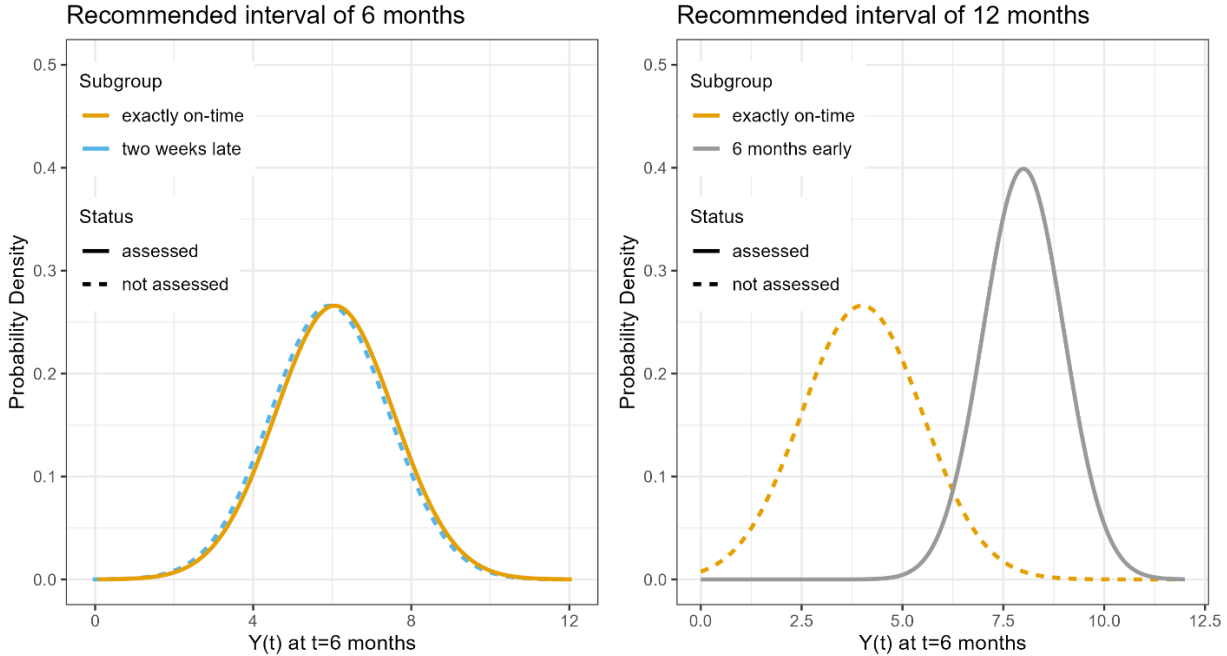

Figure 6: Illustration of our modified version of the tilt assumption for EHR data. We consider a disease outcome where higher scores indicate worse symptoms, and patients are assigned recommended visit intervals based on the outcome score measured at the current visit. In each panel, we have controlled for the observed history. In the left panel, we show that under our framework, for patients who are asked to return in 6 months, the distribution of the outcomes for patients who are assessed at exactly 6 months is the same as the distribution for patients who are non-

assessed at 6 months, but will have their assessment within a reasonable bound around 6 months, i.e. one week later. While in the right panel, where we focus on patients who are asked to return in 12 months, we show that the distribution for the early subgroup of patients, who are assessed at 6 months, is tilted with more weight on higher values of the outcome, compared to the distribution of the non-assessed patients who adhere perfectly to their 12 month recommended follow-up interval.

In summary, if some form of exponential tilting is to be applied to EHR data, it needs to reflect the fact that we do not expect a difference in outcome distributions when patients visit within a reasonable range of when they were projected to return, based on the information available at the previous visit. Thus, it is necessary to have a means of identifying when visits should plausibly occur. We employ recommended visit intervals to fulfill this role, and formulate a modified version of the exponential tilting assumption using visit categories centred around the recommended intervals.

### 5.2.1 Exponential tilting with EHR data

Previously, when we examined the IIW-GEE under AAR, we showed how we defined visit categories in the JDM study. Here, we consider the general case, where the user defines visit categories such that at any time $t$ a visit falls into exactly one category. So, we have $V_i(t) \in \mathcal{V} = \{\text{set of visit categories}\}$. We then modify Equation 4 to formulate our modified version of the exponential tilting assumption as follows, for $V_i(t) = v \in \mathcal{V}$:

$$f(Y_i(t) \mid \Delta N_i(t) = 0, \mathcal{H}_i(t)) = f(Y_i(t) \mid \Delta N_i(t) = 1, \mathcal{H}_i(t)) \frac{\exp\{q(Y_i(t), \mathcal{H}_i(t); \alpha_v)\}}{c_v(t, \mathcal{H}_i(t); \alpha_v)}$$

(6)

where $c_v(t, \mathcal{H}_i(t); \alpha_v) = E(\exp\{q(Y_i(t), \mathcal{H}_i(t); \alpha_v)\} \mid \Delta N_i(t) = 1, \mathcal{H}_i(t))$ is a normalizing constant, and we set $\alpha_{\text{in-window}} = 0$. Note that the rationale for setting $\alpha_{\text{in-window}} = 0$ is to ensure $f(Y_i(t) \mid \Delta N_i(t) = 0, \mathcal{H}_i(t)) = f(Y_i(t) \mid \Delta N_i(t) = 1, \mathcal{H}_i(t))$ for in-window visits.

The corresponding visit intensity model is as follows, for $V_i(t) = v \in \mathcal{V}$:

$$\lambda_i(t; Y_i(t), \mathcal{H}_i(t)) = \lambda_i(t; \mathcal{H}_i(t)) \frac{c_v(t, \mathcal{H}_i(t); \alpha_v)}{\exp\{q(Y_i(t), \mathcal{H}_i(t); \alpha_v)\}}$$

(7)

where the $\gamma$ parameter(s) encoded in Equation 2's formula for $\lambda_i(t; \mathcal{H}_i(t))$ is estimated through a recurrent events model (with the specific form selected by the analyst) fitted on the observed data. In addition, note that as this follows from Equation 6, we also set $\alpha_{\text{in-window}} = 0$ here.

The resulting visit intensities are then inverted to form the weights for the IIW-GEE.

## 5.3 Global sensitivity analysis of the JDM dataset
### 5.3.1 Expert elicitation

A range of plausible values for the sensitivity analysis parameters was used to vary the strength of the association between increases in DAS$_{mod}$ and visit intensity. In more detail, $\alpha_e$ represents the association between worsening health condition and the visit intensity in the early/very early time frame, and $\alpha_l$ the association between worsening health condition and the visit intensity in the late/very late time frame. Separate parameters for the early/very early and late/very late time frames were specified based on the clinical belief that early visits are often due to the patient being unwell, whereas late visits tend to be driven by a variety of factors that are not necessarily related to the health status of the patient (e.g., schedule conflicts, inclement weather hindering travel).

Clinically plausible values for the sensitivity parameters were elicited through expert opinion from a paediatric rheumatologist. Specifically, expert opinion was that the probability of visiting in the next two weeks in the absence of a scheduled visit is close to zero when there has been no increase in DAS$_{mod}$ since the last visit, increases rapidly for increases in DAS$_{mod}$ of 2-5 points, and asymptotes between 0.6 and 0.99. See Figure 4 for an example of this curve, with the asymptote occurring at 0.8. Thus, rather than $Y_i(t)$, for this analysis we work with $D_i(t) = \max(Y_i(t) - Y_i(T_{iN_i(t^-)}), 0)$, the increase (with decreases coded as zero) in DAS$_{mod}$ from the last visit before time $t$.

We experimented with different functions of $D_i(t)$ to produce the desired relationship between increasing DAS$_{mod}$ and the probability of an early visit, as outlined by our clinical expert. We found that the normal cumulative distribution function with mean=3 and standard deviation=1 produced the best fit.

The clinically plausible values for $\alpha_e$ and $\alpha_l$ were chosen as those that produced the desired 0.6-0.99 range of asymptotes, with increasing parameter values corresponding to higher asymptotes. Because we are focusing on visiting sooner rather than later, we examine early and very early visits, and find a range of plausible values for $\alpha_e$. We assume the same range for $\alpha_l$. We note that the probability of visiting also depends on the recommended interval assigned at the previous visit, as well as the visit type (early or very early). Thus, to simplify matters, we investigated the most common recommended interval categories in the very early visit classification. We identified the minimum value of $\alpha_e$, across the recommended interval categories, that allowed for a horizontal asymptote at 0.6. This gave us the lower bound for our range of plausible $\alpha_e$ values. To find the upper bound, we took the maximum value, across the recommended intervals, that allowed for a horizontal asymptote at 0.99. See Figure 5 for an illustration of this process, with examples of these curves at recommended visit intervals of 2, 6, and 12 months, and for various $\alpha_e$ values. This process led to a clinically plausible range of 4-7, but for sensitivity analysis we must begin at 0, since this corresponds to the null AAR case. Thus, $\alpha_e$ and $\alpha_l$ were varied over a grid with both parameter values ranging from 0 to 7.

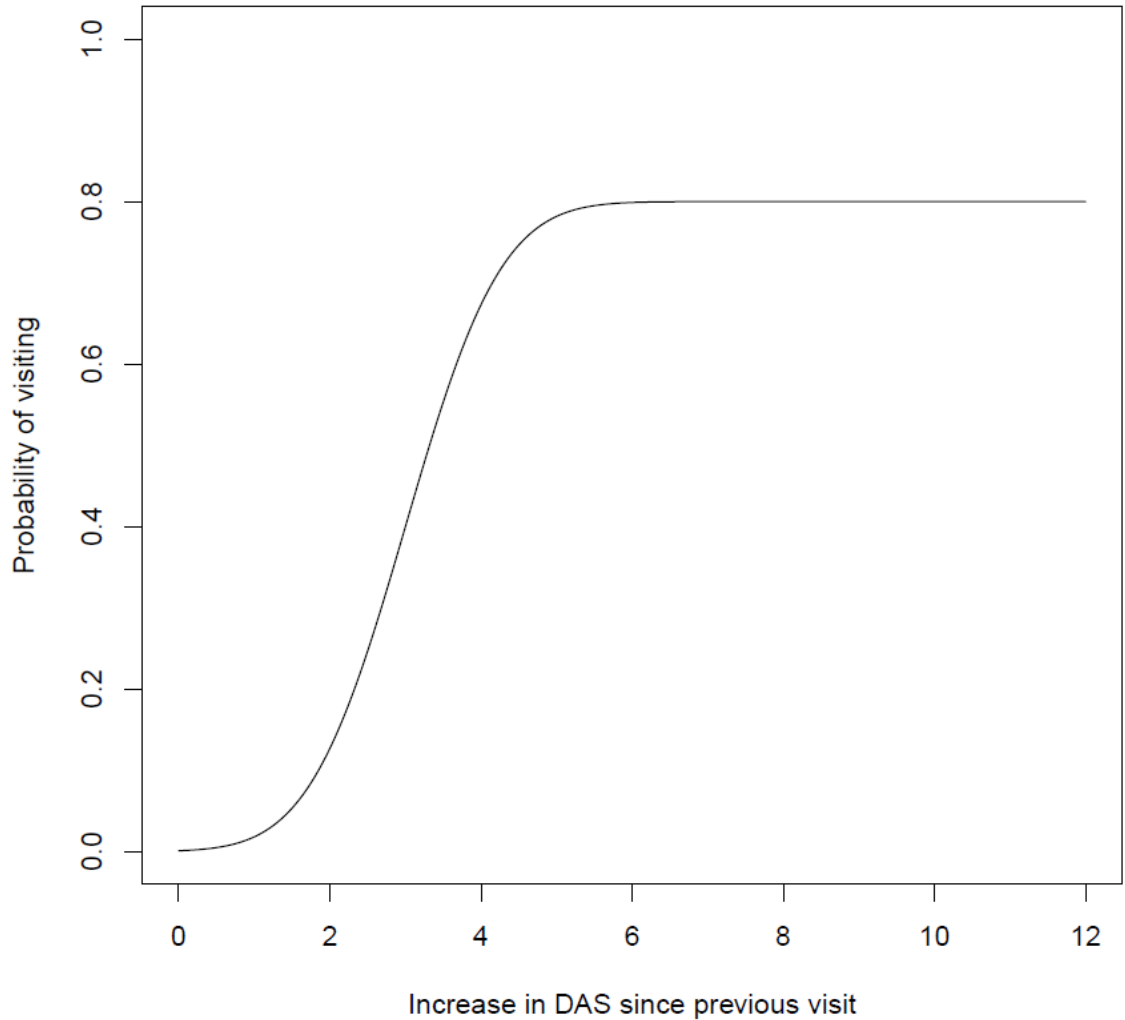

Figure 7: Clinical expert belief on the probability of visiting vs. increase in DAS since the previous visit. Note that although in practice DAS is measured on a discrete scale, for clearer visualization, this plot was generated using theoretical DAS values on a continuous scale.

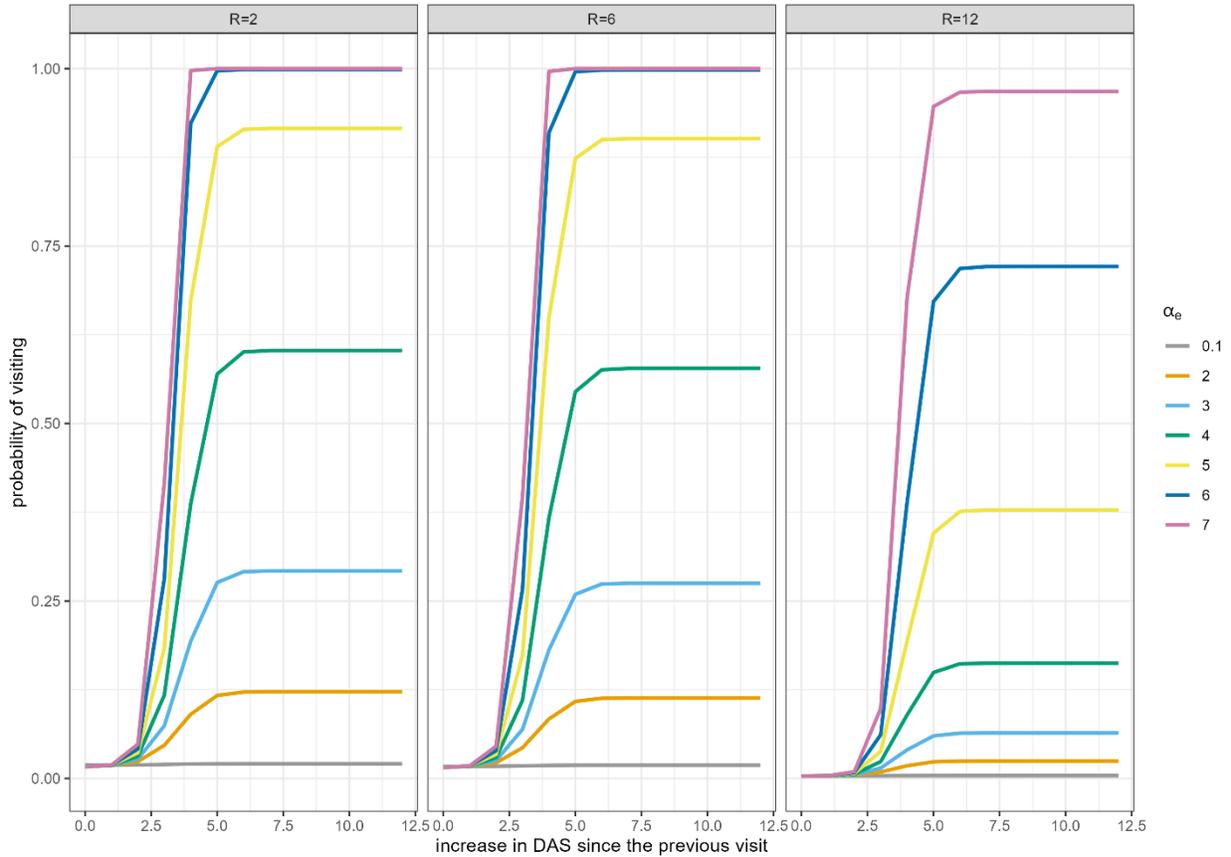

Figure 8: The calculated probability of a patient in the JDM study visiting in the next two weeks vs. the increase in DAS since the previous visit. We adjust for the recommended visit interval assigned at the previous visit and also fix the value of $\alpha_e$. We display three of the most common recommended visit intervals among very early visits, and various $\alpha_e$ values in the 0-7 range.

### 5.3.2 IIW-GEE allowing for ANAR

We then assessed the sensitivity of the disease trajectory estimated through the IIW-GEE to a specific case of ANAR, namely that outside of the in-window time frame, an increase in $DAS_{mod}$ from the previous visit was associated with a greater visit intensity.

Applying Equation (6) and setting $q(Y_i(t), \mathcal{H}_i(t); \alpha_e) = -\alpha_e \Phi(D_i(t) - 3)$ (and similar for the late visits), where $\Phi$ is the standard normal cumulative distribution function, we formulate the visit intensity model as follows:

$$\lambda_i(t; R_{iN_i(t^-)}, T_{iN_i(t^-)}, D_i(t))$$

$$= \begin{cases} \lambda_i(t; R_{iN_i(t^-)}, T_{iN_i(t^-)}) \dfrac{E(\exp(-\alpha_e \Phi(D_i(t) - 3)) \mid R_{iN_i(t^-)}, \Delta N_i(t) = 1)}{\exp(-\alpha_e \Phi(D_i(t) - 3))}, & V_i(t) \in \{\text{early, very early}\} \\[1em] \lambda_i(t; R_{iN_i(t^-)}, T_{iN_i(t^-)}) \dfrac{E(\exp(-\alpha_l \Phi(D_i(t) - 3)) \mid R_{iN_i(t^-)}, \Delta N_i(t) = 1)}{\exp(-\alpha_l \Phi(D_i(t) - 3))}, & V_i(t) \in \{\text{late, very late}\} \\[1em] \lambda_i(t; R_{iN_i(t^-)}, T_{iN_i(t^-)}), & V_i(t) \in \{\text{in-window}\} \end{cases}$$

(9)

where we note that the expression for the visit intensity for in-window visits arises from setting $\alpha_{\text{in-window}} = 0$.

## 5.4 Results: sensitivity analysis

Figure 2 shows the disease activity trajectories estimated by the unweighted GEE and three different IIW-GEEs. The IIW-GEE assuming AAR corresponds to the $\alpha_e = 0$ curve. In all models, the estimated mean disease activity drops in the first few years after diagnosis. The IIW-GEE curve estimate retains the same shape but is lower as $\alpha_e$ increases from 0 to 7. The heatmap in Figure 9 shows that the estimated trajectories are more sensitive to $\alpha_e$ than to $\alpha_l$: for a given $\alpha_l$ value, the AUC decreases by ~2.4 as $\alpha_e$ increases from 0 to 7, but decreases by ~0.2 as $\alpha_l$ increases from 0 to 7 (for a given $\alpha_e$ value). For example, setting $\alpha_l = 0.5$, the AUC decreases from 19.2 to 16.8 as $\alpha_e$ increases from 0 to 7. Conversely, setting $\alpha_e = 0.5$, the AUC decreases from 18.7 to 18.5 as $\alpha_l$ increases from 0 to 7. Within the clinically plausible range of 4 to 7, the changes in AUC are relatively small, but the values are markedly different from the standard $\alpha_e = 0$, $\alpha_l = 0$ curve; for example, the AUC for the standard curve is 19.2, while the AUC ranges from 17.2 to 16.9 over the clinically plausible range of $\alpha_e$ (setting $\alpha_l = 0$). Analysis using complete cases suggests similar results in terms of these trends (see Supplementary Figures S1 and S2).

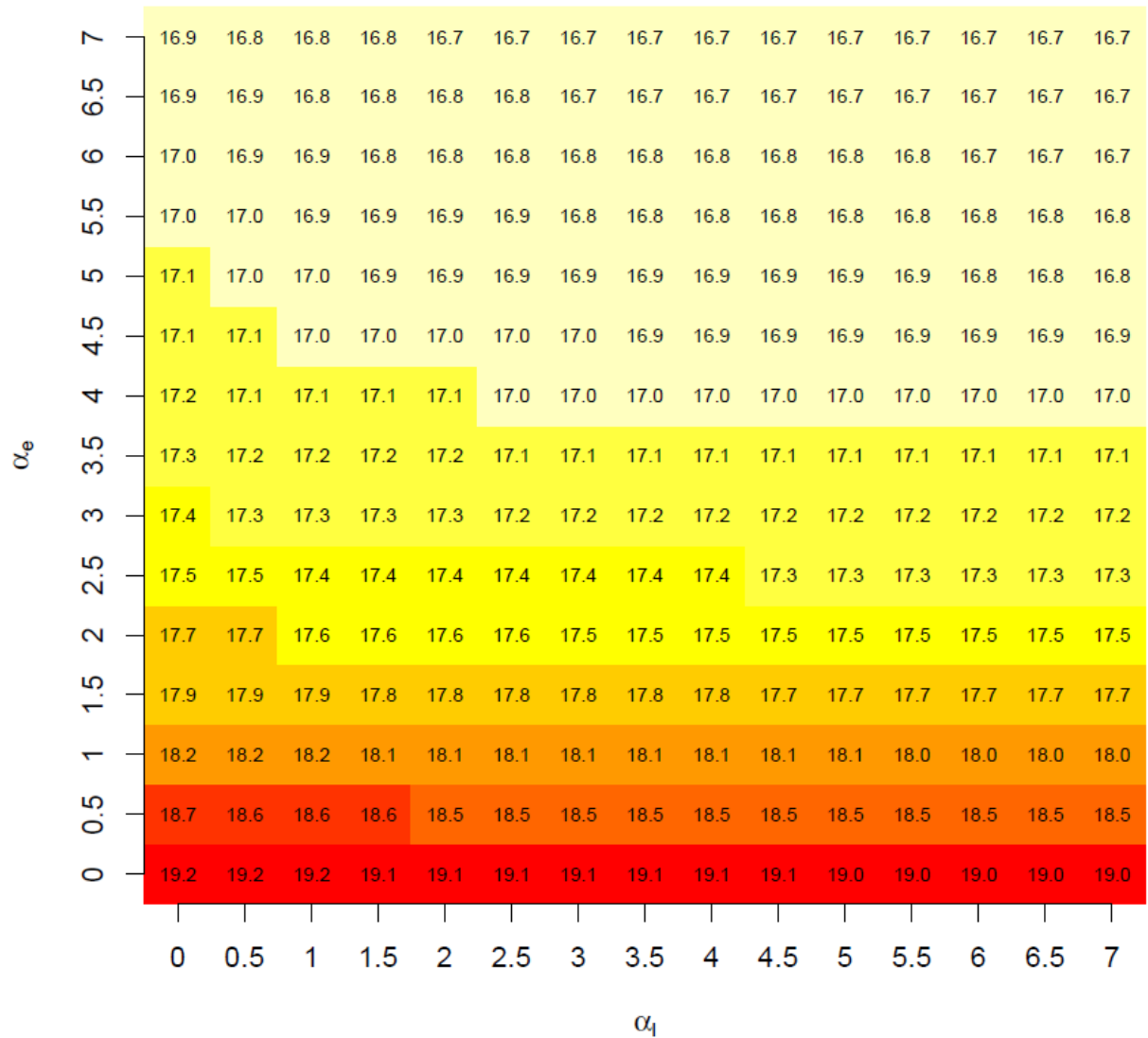

Figure 9: Area Under the Curve (AUC) for the IIW-GEE over the entire grid of values for $\alpha_e$ and $\alpha_l$. The AUC for the unweighted GEE is 19.9. Note: AUC values are rounded to one decimal place.

## 6 DISCUSSION

In this paper, we proposed an approach to modelling longitudinal data subject to irregular observation that leverages recommended visit intervals to examine the irregular visit mechanism and assess the potential impact of ANAR. See Appendix D in the Supporting Information for the tutorial on how to code the sensitivity analysis in R.

Figure 1 shows that in our example dataset, there were highly irregular visit times. However, we found that the vast majority (78%) of the variability in visit times could be explained by the physician-recommended intervals between visits. In addition, Figure 2 and Figure 3 show overall close agreement between observed and recommended visit intervals.

Moreover, we demonstrated how extracting the recommended visit intervals makes global sensitivity analysis possible; in order to investigate the potential impact of ANAR, we need to first formulate a model representing the ANAR mechanism. Recommended visit intervals fulfil this vital role; they allow us to define visit categories so that we can model a specific case of ANAR where an out-of-window visit is more likely if an exacerbation of the disease has occurred. The sensitivity analysis showed that there was a potential impact of this specific type of ANAR on the disease trajectory curves estimated by the IIW-GEE. Specifically, Figure 6 shows that the IIW-GEE curves were sensitive to increasing the strength of the association between worsening health condition and the visit intensity in the early/very early time frame but were robust to increasing the strength of the association in the late/very late time frame. This aligns with our intuition, as patients may come in for unexpected visits earlier than recommended because they need more urgent care (e.g., due to a disease flare), and patients likely visit late for many different reasons – for example, other life priorities that are unrelated to their health status.

Furthermore, in alignment with previous work,[1,2,5] once we account for the irregularity in visit times, the estimated mean disease trajectory curve drops. This corresponds to our intuition that the standard GEE model overestimates the burden of the disease, since there is an overrepresentation of measurements on sicker patients.

A limitation of the proposed work is that our approach is dependent on the visit categories that were used to categorize the visits. In our example we relied on a combination of clinical expertise and judgement calls by the analyst. Thus, there will always be some subjective decisions involved. In addition, there are sudden changes in the intensities due to our categorization of the visits, where smoother transitions are more plausible. A kernel weighting approach might be interesting to explore in future work.

Another challenge is that EHR datasets often feature missing data. We have performed multiple imputation assuming missing at random, but we have not examined the sensitivity of the results to missing not at random.

Furthermore, although we modelled the longitudinal disease trajectory using IIW-GEEs, there are other approaches available for analyzing irregularly observed longitudinal data. Pullenayegum and Lim[1] provide an extensive review of these methods, and outline the conditions under which each method is applicable. Briefly, we will note that semiparametric joint models are appropriate for special cases of AAR and special cases of ANAR, which are distinct from the assessment process encountered in our example. Fully parametric joint models offer an alternative option; however, this approach is not yet well-established and there is limited software available to implement these models. Future work will leverage the recommended visit intervals to formulate a fully parametric joint model and may also explore a Bayesian joint model.

The proposed sensitivity analysis does not produce a single final answer, in terms of which IIW-GEE model to rely upon. In our example, we had fairly consistent results across the range of plausible $\alpha_e$ and $\alpha_l$ values, so any of the IIW-GEEs allowing for ANAR could potentially be presented as a reasonable final model, but this will not necessarily be the case in other studies. In the absence of a sensitivity analysis, researchers simply proceed with the IIW-GEE that assumes AAR and ignore the potential impact of ANAR. However, we argue that examining the sensitivity of the results to assumptions about the irregular visit mechanism should become a mandatory element of reporting, even if this causes the interpretation of the results to be less straightforward. Moreover, this work could be extended to a Bayesian framework by developing a joint model for the outcome and visit processes and eliciting expert opinion to specify distributions for the sensitivity analysis association parameters. This would produce a single posterior distribution of the desired estimand in place of the multiple estimates returned by the frequentist approach to sensitivity analysis.

We also note that in studies going forward, it will be possible to extract information on the recommended visit intervals automatically, thus improving the feasibility of the proposed methods in this paper. In more detail, while we relied on manual chart reviews and examination of physicians' notes to obtain the recommended intervals, these now a field in modern EHRs, as orders (e.g. follow-up in 6 weeks) must be written and filled in order to schedule subsequent visits.

In summary, our results highlight the strong ability of recommended gap times to enhance the rigour of longitudinal data analyses featuring irregular visit times, and thus, we advocate that extraction and analysis of recommended gap times should become a routine component of study design and reporting.

# Supporting information

**Summary**

**Appendix A**: Supplemental Figure S1: flowchart describing the sample size and final number of observations in the dataset.

**Appendix B**: Supplemental Tables 1 and 2.

**Appendix C**: Supplemental Figures S2 and S3 showing results under complete case analysis.

**Appendix D**: R tutorial on how to use recommended visits intervals to fit IIW-GEEs under AAR and ANAR.

# APPENDIX A

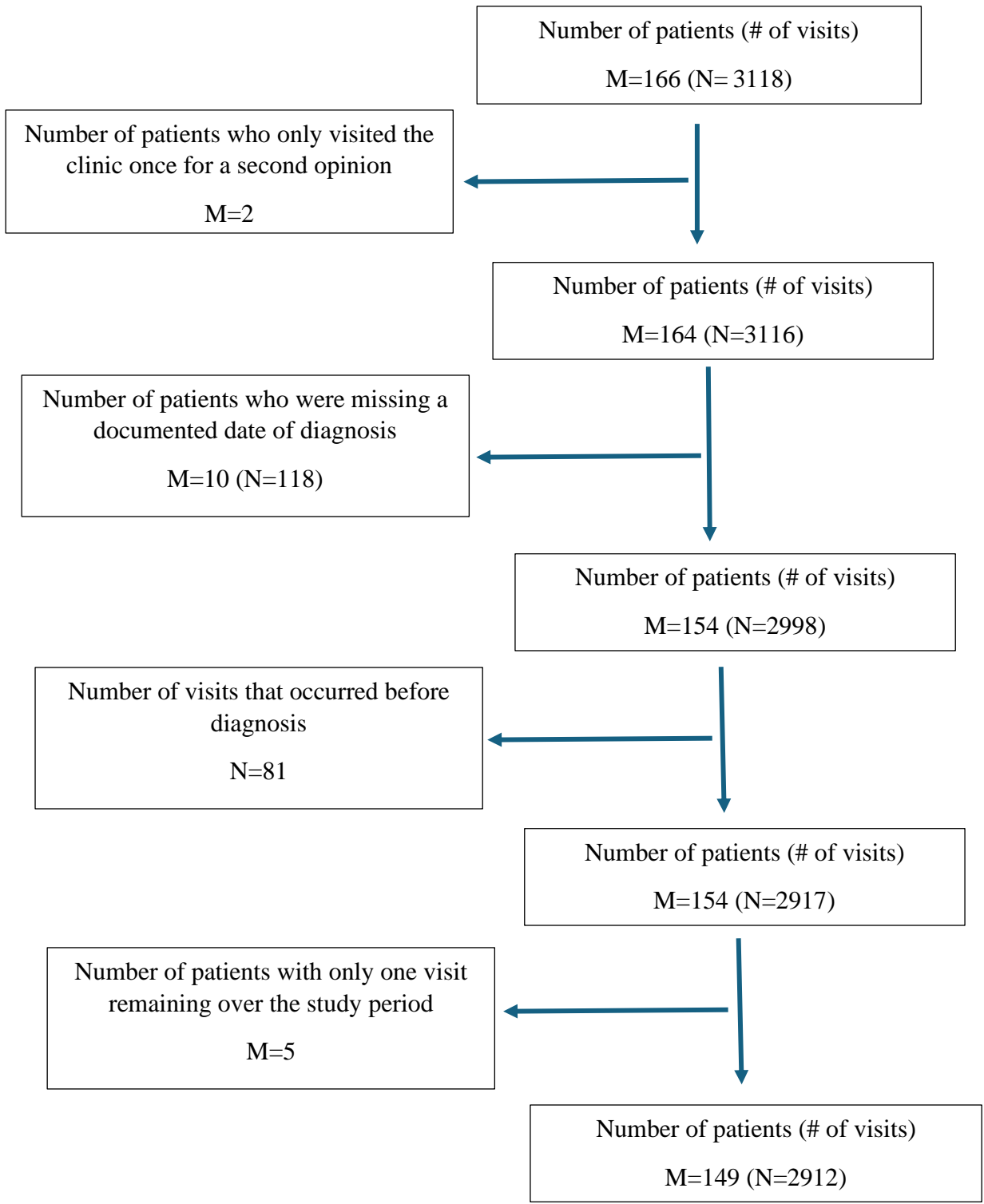

Figure S1: Flowchart describing the sample size and final number of observations in the dataset.

# APPENDIX B

**TABLE 1.** Individual-level characteristics of a cohort of 149 patients with juvenile dermatomyositis, with a grand total of 2912 visits observed over the study period, from 2000 to 2018.

| Patient-Level Characteristics | N with % or median [interquartile range] |
|---|---|
| Sex, Female | 97 (65.1%) |
| Age at diagnosis (years) | 7.8 [4.8, 11.7] |
| Race/Ethnicity | |
|   Asian | 28 (18.8%) |
|   Black | 7 (4.7%) |
|   Hispanic or Latino | 3 (2.0%) |
|   White | 75 (50.3%) |
|   Unknown | 32 (21.5%) |
|   Other | 4 (2.7%) |
| Duration of follow-up (years) | 6.9 [3.4, 10.9] |
| Number of visits | 18 [10, 26] |

Note: duration of follow-up is measured from date of diagnosis, even if the patient was diagnosed before the study period began. The number of visits refers to visits occurring over the study period only.

**TABLE 2.** Visit-level characteristics of a cohort of 149 patients with juvenile dermatomyositis, with a grand total of 2912 visits observed over the study period, from 2000 to 2018.

| Visit-Level Characteristics | Median [interquartile range]; minimum to maximum |
|---|---|
| Observed visit interval (months) | 2.8 [1.4, 4.3]; 1 day to 3.4 years |
| Recommended visit interval (months) | 3 [1.4, 4.0]; 3 days to 14 months |
| Ratio of observed to recommended visit intervals | 1 [0.92, 1.1]; 0.07 to 40 |
| Difference between observed and recommended intervals (months) | 0 [-0.16, 0.42]; -11.1 to 28.5 |
| Modified DAS | 3 [1, 5]; 0 to 12 |

Note: 62 visit intervals were censored at the end of the study period and were therefore excluded when calculating the summary statistics for the observed visit intervals, and the ratio and difference between observed and recommended visit intervals.

# APPENDIX C

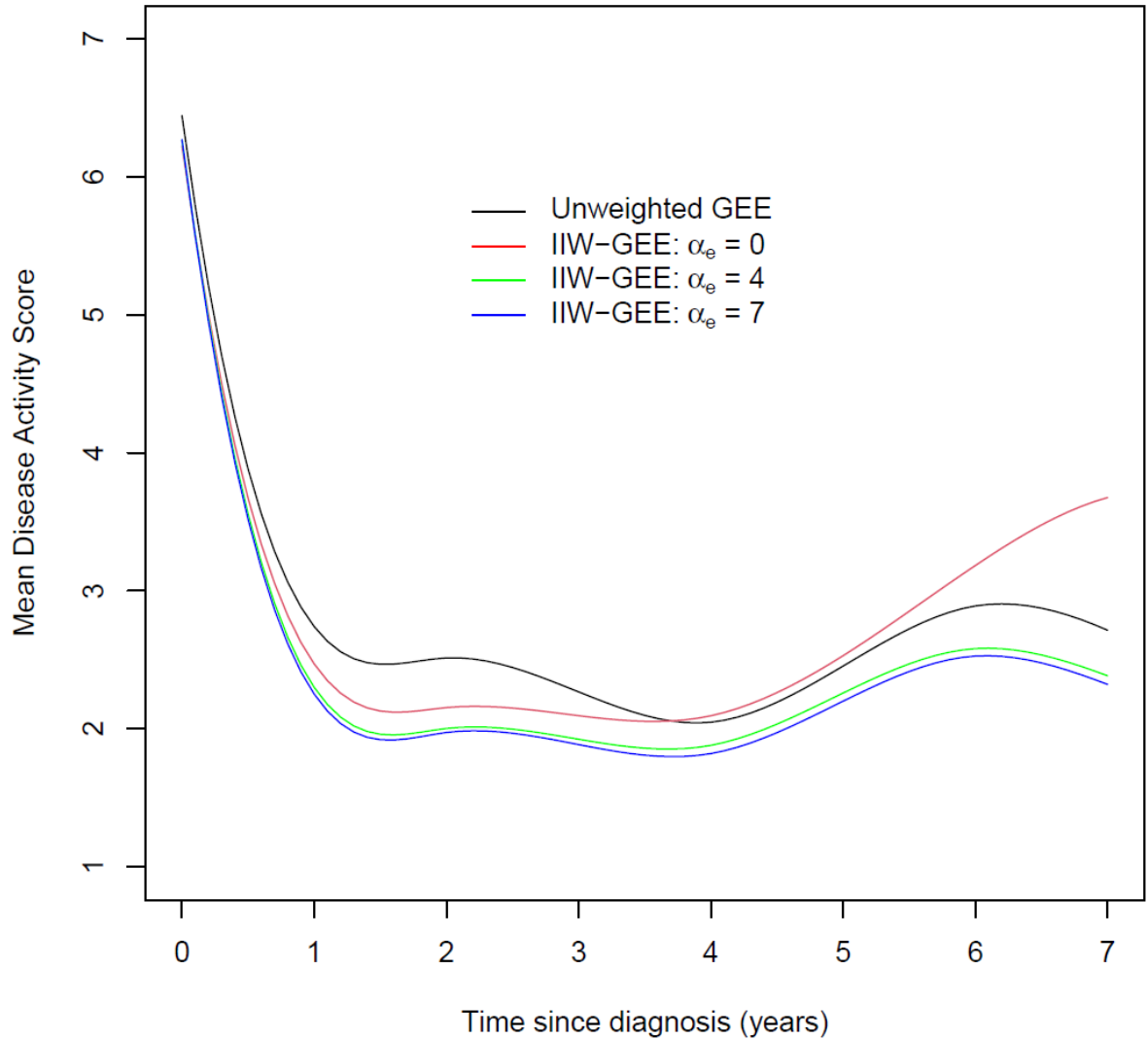

Figure S2: Fitted expected disease activity as a function of time using unweighted and inverse-intensity weighted (IIW) GEEs under complete case analysis. The association parameter quantifying the strength of the association between the hazard of visiting earlier than recommended and worsening health condition, $\alpha_e$, is shown at 3 example values specified through the sensitivity analysis. The $\alpha_l$ parameter representing the strength of the association between the hazard of visiting later than recommended and worsening condition is set to be 0. The red curve (with $\alpha_e = \alpha_l = 0$) represents the IIW-GEE assuming AAR.

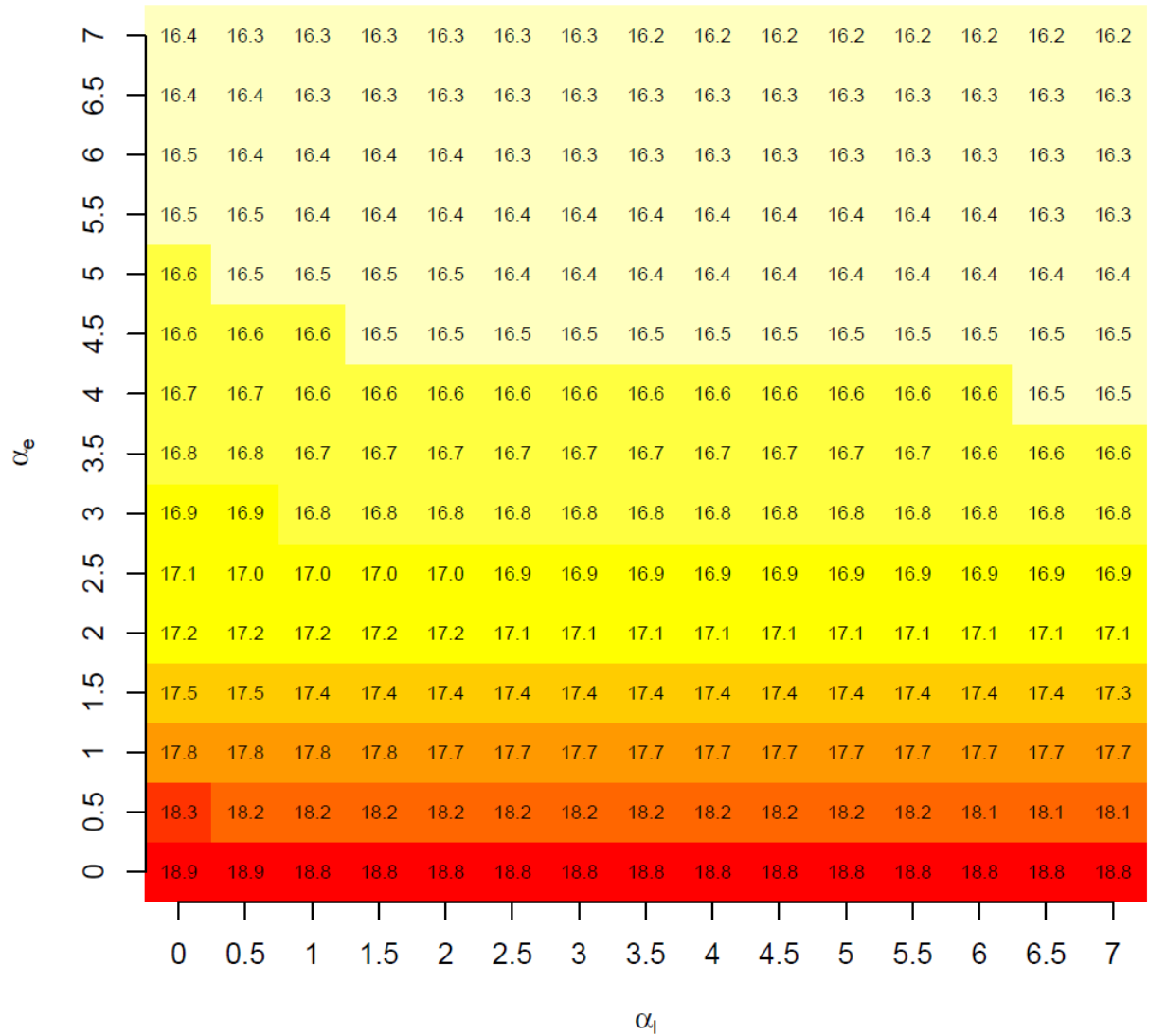

Figure S3: Area Under the Curve (AUC) for the IIW-GEE under complete case analysis over the entire grid of values for $\alpha_e$ and $\alpha_l$. The AUC for the unweighted GEE is 19.0. Note: AUC values are rounded to one decimal place.

# Appendix D
# Tutorial: how to use recommended visits intervals to fit IIW-GEEs under AAR and ANAR

**Goal:** The goal of this tutorial is to provide guidance on how to use recommended visit intervals to fit IIW-GEEs under both assessment at random (AAR) and a specific case of assessment not at random (ANAR). We also show how to assess sensitivity of results to ANAR through plots. All codes are in `R`.

**Note:** In the main body of the paper, we imputed missing values of the recommended visit intervals and the disease outcome, but for simplicity, in this tutorial we present a complete case analysis approach.

## Example dataset

In this tutorial, we illustrate the implementation of our proposed method using a dataset from a clinic-based cohort study on juvenile dermatomyositis. In our example, the disease outcome is a disease activity score (DAS) ranging from 0 to 12.

Below we show what a dataset to be used in this analysis might look like. Note that this is fake data, generated for the purpose of illustration.

```
head(fakedat)
```

```
## # A tibble: 6 x 7
##      id date       time_since_dx   DAS      S censor      R
##   <int> <date>             <dbl> <int>  <dbl>  <int>  <dbl>
## 1     1 2009-05-13        0.0383    10  0.690      0  0.460
## 2     1 2009-06-03        0.0958    10  0.460      0  0.460
## 3     1 2009-06-17        0.134      7  1.38       0  2
## 4     1 2009-07-29        0.249     NA  2.30       0  2
## 5     1 2009-10-07        0.441      5  4.14       0 NA
## 6     1 2010-02-10        0.786      3  4.60       0  2
```

## Data set-up

- `id`: numeric subject ID
- `date`: date of the clinic visit
- `time_since_dx`: time since diagonsis (in years)
- `DAS`: the disease activity measured at the given visit
- `S`: the time interval between visits (in months)
- `censor`: dummy variable indicating whether the interval `S` is censored by the end of the study period
- `R`: the recommended visit interval (in months)

Example measurements from fake dataset:

$S_{11} \approx 0.690$ months=3 weeks elapsed between the first and second visit for subject 1. $R_{11} \approx 0.460$ months=2 weeks, so the physician recommends that patient 1 returns for their second visit in 2 weeks. Thus, we can see that the patient returns 1 week later than recommended for their second visit.

Notes:

1. `censor=1` only for the final observation for individuals who have their follow-up cut off by the end of the study. If follow-up has truly ended (e.g. patient transferred to adult care/ different clinic), then the final observation for that individual will have `censor=0`.

2. We converted all actual and recommended visit intervals to the month time scale (assuming that there is on average 4.345 weeks per month, and 30.417 days per month) because that was the most popular unit of measurement in this study. The most suitable time unit will depend on the specific application.

In this tutorial, we use the following `R` packages: `tidyverse` for data manipulation (Wickham et al. (2019)), and `survival` (Therneau and Grambsch (2000), Therneau (2021)), `geepack` (Højsgaard, Halekoh, and Yan (2006), Yan and Fine (2004), Yan (2002)), and `splines` (R Core Team (2021)) for model-fitting.

## Preparing the dataset for analysis

We go through the 3 steps required to prepare our dataset for analysis.

### 1. Create visit window indicator variables

First, we create indicator variables `very_early`, `early`, `late`, `very_late`, and `in_window`, classifying visits based on the thresholds discussed in the main body of the paper.

### 2. Calculate the time-at-risk of each event type (each type of visit)

Next, we calculate the time-at-risk of each type of event (i.e. each type of visit), as we need this to estimate the intensities for each type of event. We append 5 additional columns to the dataframe: `very_early_time`, `early_time`, `late_time`, `very_late_time`, and `in_window_time`. We assume a gap-time scale, where the time index is reset to zero after each visit.

**Examples**:

We go through how to calculate the time-at-risk for the observations shown for `fakedat`. Our calculations are based on the clinical thresholds decided upon for this dataset. Thus, this is specific to this dataset and this disease, and decisions will vary depending on the clinical thresholds chosen for a given problem.

At the end of both the first and second visits in the example dataset, a recommended interval of 2 weeks is assigned. In this case, it is impossible for the visit to be very early or early, since to be very early, the visit has to occur more than 1 month earlier than recommended, and to be early, the visit has to occur more than 2 weeks earlier than recommended. So, visits with these recommended intervals can only be in-window, late, or very late. In this example, the second and third visits are both in-window. The second visit contributes its full interval of 3 weeks to the time-at-risk of an in-window visit, and does not contribute any time-at-risk to the other 4 visit types. The third visit similarly contributes its full interval of 2 weeks to the time-at-risk of an in-window visit.

A recommended visit interval of 2 months is assigned at the end of the third visit, but the patient ends up visiting 1.38 months later. This indicates an early visit, and this observation contributes 1 month to the time-at-risk of a very early visit, and 0.38 months (the time remaining) to the time-at-risk of an early visit.

At the end of the fourth visit, a recommended interval of 2 months is given, but the patient ends up visiting 2.3 months later. This indicates an in-window visit, where the first 1 month is time-at-risk of a very early visit, the next 0.5 months is time-at-risk of an early visit, and the remainder (0.8 months) is time-at-risk of an in-window visit.

The recommended interval assigned during the fifth visit is missing, so it is impossible to calculate the time-at-risk of each type of visit.

Finally, for the last observation, the recommended interval is 2 months, but the patient ends up visiting 4.6 months later. This indicates a very late visit, so this observation contributes time-at-risk to all five types of visits: 1 month for very early, 0.5 months for early, 1.5 months for in-window, 1 month for late, and the remainder (0.6 months) for very late.

Note: the time at risk should never be zero - if an observation does not contribute any time-at-risk to a given visit type, set the contribution to `NA` rather than zero. This will be important when running the visit intensity models later on, as these survival models will not accept survival times of 0.

### 3. Calculate change in disease outcome

The last step in preparing the dataset for analysis is to create the variable `diff_das`, which encodes $\Delta Y_i(t)$.

We also add `diff_das_plus`, which encodes $D_i(t)$, as defined in the main body of the paper.

Going forward into the next part of the tutorial on analyzing the data, we will be working with the real dataset, which is called `mydat`.

## Visit intensity model assuming AAR

We will now fit the IIW-GEE assuming AAR. First, we have to formulate a model for the visit intensity, to estimate the intensities that will be later used in weighting.

Initially, we intended to fit separate exponential regressions for each category of recommended interval and each of the five visit types; a fully stratified approach. However, in this dataset, there are some recommended interval types that are relatively rare, so the sample size in these cases was too small to provide a good model fit. Thus, in order to use all of the data, we instead use a smooth function of the recommended visit interval as the sole predictor in the exponential survival model for each visit type.

Note: We fit the out-of-window visit intensity models only on observations for which there is an observed `diff_das`. This is done to ensure that we are using the same data when we fit the models in the sensitivity analysis that allow for ANAR.

The model-fitting codes are shown below:

```
early_surv<- mydat %>%
  filter(!is.na(diff_das)) %>%
  survreg(Surv(early_time, early)~ splines::bs(R),
          dist="exponential", data= ., na.action = na.exclude)

very_early_surv<- mydat %>%
  filter(!is.na(diff_das)) %>%
  survreg(Surv(very_early_time, very_early)~ splines::bs(R),
          dist="exponential", data= ., na.action = na.exclude)

late_surv<- mydat %>%
  filter(!is.na(diff_das)) %>%
  survreg(Surv(late_time, late)~ splines::bs(R),
          dist="exponential", data= ., na.action = na.exclude)

very_late_surv<- mydat %>%
  filter(!is.na(diff_das)) %>%
  survreg(Surv(very_late_time, very_late)~ splines::bs(R),
          dist="exponential", data= ., na.action = na.exclude)

in_window_surv<- mydat %>%
  survreg(Surv(in_window_time, in_window)~ pspline(R, df=2),
          dist="exponential", data= ., na.action = na.exclude)
```

Now we use the models to extract the estimated visit intensities.

```
#first find the indices of the observations that are not missing diff_das
not_missing<- which(!is.na(mydat$diff_das))
#now calculate the intensities for these observations
calc_intensity<- function(survmodel, outofwindow="YES"){
  if (outofwindow=="YES"){
    #padding vector with NA's ensures length(intvect)=nrow(mydat)
    intvect<- rep(NA, nrow(mydat))
    intvect[not_missing] <- exp(-predict(survmodel, type="lp"))
    intvect
  }
  #in-window visit intensities are unchanged under ANAR,
```

```r
  #no need to worry about observations with missing diff_das
  else{
    intvect<- exp(-predict(survmodel, type="lp"))
    intvect
  }
}

##add the intensies calculated for each model to the dataframe
mydat<- mydat %>% add_column(int_early=calc_intensity(early_surv),
                             int_very_early=calc_intensity(very_early_surv),
                             int_late= calc_intensity(late_surv),
                             int_very_late=calc_intensity(very_late_surv),
                             int_in_window=calc_intensity(in_window_surv,
                                                          outofwindow = "NO"))

#the actual visit intensity we will use for the weights depends on what type of visit it is.
mydat <- mydat %>%
  mutate(int_aar = case_when(early==1 ~ int_early,
                             very_early==1 ~ int_very_early,
                             late==1 ~ int_late,
                             very_late==1 ~ int_very_late,
                             in_window==1 ~ int_in_window,
                             TRUE ~ NA_real_))
```

## Calculating inverse-intensity weights

Now, we can begin to calculate the inverse-intensity weight for each observation.

```r
mydat <- mydat %>%
  mutate(raw_weight = 1/int_aar)
```

After taking the inverse of the visit intensity, two further adjustments to the weights were required. Since the weights were initially aligned with the recommended gap provided by the physician at a given visit j, within each individual each weight had to first be shifted up by one time point to be aligned instead with the DAS for the following visit:

```r
mydat<- mydat %>%
  group_by(id) %>%
  mutate(lag_weight =  dplyr::lag(raw_weight, n = 1, default = NA))
```

Then, a weight of one was assigned to the first visit for each individual:

```r
mydat<- mydat %>%
  group_by(id) %>%
  mutate(myweight = case_when( date==min(date) ~ 1,
                               date!=min(date) ~ lag_weight))
```

## Fitting GEEs: unweighted and IIW-GEE assuming AAR

Now, we have the weights, so we can estimate the IIW-GEE assuming AAR. But first, let's fit the unweighted GEE. The GEEs regressed the disease outcome, DAS, onto a cubic smoothing spline function of the time since diagnosis.

```r
unweighted_gee<- mydat %>%
  filter(!is.na(myweight)) %>%
  geeglm(DAS ~ pspline(time_since_dx, df=3), id=id,
         family=gaussian, corstr="independence",
         data = .)
```

Now the weighted GEE:

```
weighted_gee<- mydat %>%
  geeglm(DAS ~ pspline(time_since_dx, df=3), id=id,
         family=gaussian, corstr="independence",
         data = .,
         weights= myweight)
```

To visualize the differences in the estimates produced by the two models, we can create a plot comparing the estimated mean DAS trajectory over time under the unweighted GEE to the weighted GEE assuming AAR.

```
time_plot <-  seq(0, 7, by = 0.1)
weighted_das <- predict(weighted_gee,newdata=data.frame(time_since_dx=time_plot))
unweighted_das <- predict(unweighted_gee,newdata=data.frame(time_since_dx=time_plot))

plot(time_plot ,unweighted_das,type="l", xlim=c(0,7),
     ylim=c(1,7),
     xlab="Time",ylab="Mean Disease Activity Score", main="")
lines(time_plot,weighted_das,col=2)
legend (2, 6,legend=c("Unweighted","Inverse-intensity weighted"),col=1:2,bty="n",lty=1)
```

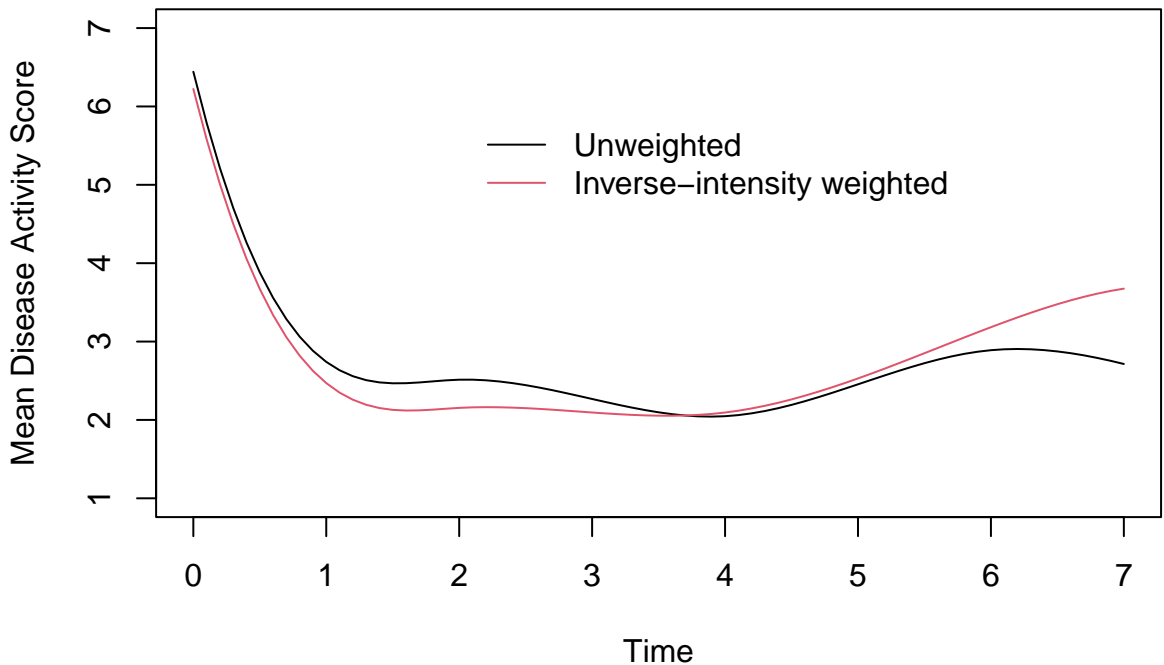

```
q_fn<- function(x){
  pnorm(q=x, mean=3, sd=1)
}
```

**Fitting IIW-GEEs allowing for ANAR**

To fit IIW-GEEs allowing for ANAR, we must calculate $c_v$, as defined in the main of the paper. To do this, we fit linear regression models to the data for out-of-window visits, with $\exp\bigl(-\alpha\Phi\bigl(D_i(t)-3\bigr)\bigr)$ as the response, and a spline function of the recommended visit interval as the sole predictor.

We now show a function that fits our proposed IIW-GEE model allowing for ANAR, with the function arguments

$\alpha_e$ and $\alpha_l$:

```r
# to make the code tidier, first code a helper function
# (this is related to the tilting function discussed in the main text)
q_fn<- function(x){
  pnorm(q=x, mean=3, sd=1)
}

GEE_ANAR<- function(alpha_e=0.1, alpha_l=0.1){
  # linear regressions for exp(-alpha*q_fn(diff_das_plus))----
  early_lm<- mydat %>%
    lm(exp(-alpha_e*q_fn(diff_das_plus)) ~ splines::bs(R),
       data= ., na.action = na.exclude)
  
  late_lm<- mydat %>%
    lm(exp(-alpha_l*q_fn(diff_das_plus)) ~ splines::bs(R),
       data= ., na.action = na.exclude)
  
  # calculating c using the above lm models ----------------------------
  mydat<- mydat %>% add_column(c_early=1/(ifelse(mydat$early==1,
                                                 fitted(early_lm), NA)),
                               c_very_early=1/(ifelse(mydat$very_early==1,
                                                      fitted(early_lm), NA)),
                               c_late= 1/(ifelse(mydat$late==1,
                                                 fitted(late_lm), NA)),
                               c_very_late=1/(ifelse(mydat$very_late==1,
                                                     fitted(late_lm), NA))
  )
  
  
  # now calculate intensities for out-of-window visits, allowing for ANAR--------
  mydat<- mydat %>%
    mutate(int_anar = case_when(early==1 ~ int_aar*(1/c_early)*exp(alpha_e*q_fn(diff_das_plus)),
                                very_early==1 ~ int_aar*(1/c_very_early)*exp(alpha_e*q_fn(diff_das_plus)),
                                late==1 ~ int_aar*(1/c_late)*exp(alpha_l*q_fn(diff_das_plus)),
                                very_late==1 ~ int_aar*(1/c_very_late)*exp(alpha_l*q_fn(diff_das_plus)),
                                in_window==1 ~ int_aar, #in-window visit intensities are unchanged
                                TRUE ~ NA_real_))
  
  
  # calculate new weights, allowing for ANAR ---------------------------------
  mydat <- mydat %>%
    mutate(raw_weight_anar = 1/int_anar)
  
  #align weights properly with disease outcome
  mydat<- mydat %>%
    group_by(id) %>%
    mutate(lag_weight_anar =  dplyr::lag(raw_weight_anar, n = 1, default = NA))
  
  #assign a weight of one to the first observation of each individual.
  mydat<- mydat %>%
    group_by(id) %>%
    mutate(myweight_anar = case_when( date==min(date) ~ 1,
                                      date!=min(date) ~ lag_weight_anar))
  
  # weighted GEE allowing for ANAR ----------------------------------------
  weighted_gee_anar<- mydat %>%
```

```
    geeglm(DAS ~ pspline(time_since_dx, df=3), id=id,
           family=gaussian, corstr="independence",
           data = .,
           weights= myweight_anar)
  return(weighted_gee_anar)
}
```

**Codes to create figures for sensitivity analysis**

Now, we provide functions to calculate the AUC and obtain predicted values for mean DAS to be used in plotting. Note that we set the default value of the `timerange` argument to be 7 years, as this was the (rounded) median duration of follow-up in our study. This can easily be adjusted depending on the application. The `increment` argument also can be tuned, either with finer or coarser increments as appropriate.

```
calculate_AUC_GEE<- function(geemod=unweighted_gee, timerange=7, increment=0.007){
  xnew <-  seq(0, timerange, by = increment)
  n= length(xnew)
  dx<- increment
  f <- predict(geemod,newdata=data.frame(time_since_dx=xnew))
  w <- rep(dx,n);w[1]<-w[n] <- w[2]/2 ## trapezoidal weights
  auc_calc<-sum(w*f) ## integral
  return(auc_calc)
}

calculate_mean_DAS<- function(geemod=unweighted_gee, timerange=7, increment=0.1){
  xnew <-  seq(0, timerange, by = increment)
  ypred <- predict(geemod,newdata=data.frame(time_since_dx=xnew))
  return(ypred)
}
```

Code to create heatmap figure:

```
alpha_e_vect<- seq(0,7, by=0.5)
alpha_l_vect<- seq(0,7, by=0.5)
grid_dim<- length(alpha_e_vect)
# produce matrix of AUCs over range of alpha
matrix_of_aucs<- matrix(0, nrow=grid_dim, ncol=grid_dim)
for (i in 1:grid_dim){
  for (j in 1:grid_dim){
    matrix_of_aucs[i,j]<- calculate_AUC_GEE(geemod=GEE_ANAR(alpha_e =  alpha_e_vect[i],
                                                alpha_l=alpha_l_vect[j]))
  }
}
# for ease of interpretation, round the AUC values to the nearest decimal place before plotting
matrix_of_aucs_rounded<- matrix(0, nrow=grid_dim, ncol=grid_dim)
for (i in 1:grid_dim){
  for (j in 1:grid_dim){
    matrix_of_aucs_rounded[i,j]<- round(matrix_of_aucs[i,j],1)
  }
}
colnames(matrix_of_aucs_rounded)<- alpha_l_vect
rownames(matrix_of_aucs_rounded)<- alpha_e_vect

# create the heatmap plot. feel free to experiment with different colour palettes!
image(1:ncol(matrix_of_aucs_rounded), 1:nrow(matrix_of_aucs_rounded),
      t(matrix_of_aucs_rounded), col = rev(heat.colors(8)),
      axes = FALSE, ylab=expression(alpha[e]),
```

```
      xlab=expression(alpha[l]))
axis(1, 1:ncol(matrix_of_aucs_rounded), colnames(matrix_of_aucs_rounded))
axis(2, 1:nrow(matrix_of_aucs_rounded), rownames(matrix_of_aucs_rounded))
for (x in 1:ncol(matrix_of_aucs_rounded))
  for (y in 1:nrow(matrix_of_aucs_rounded))
    text(x, y, format(matrix_of_aucs_rounded[y,x], nsmall=1), cex=0.7)
```

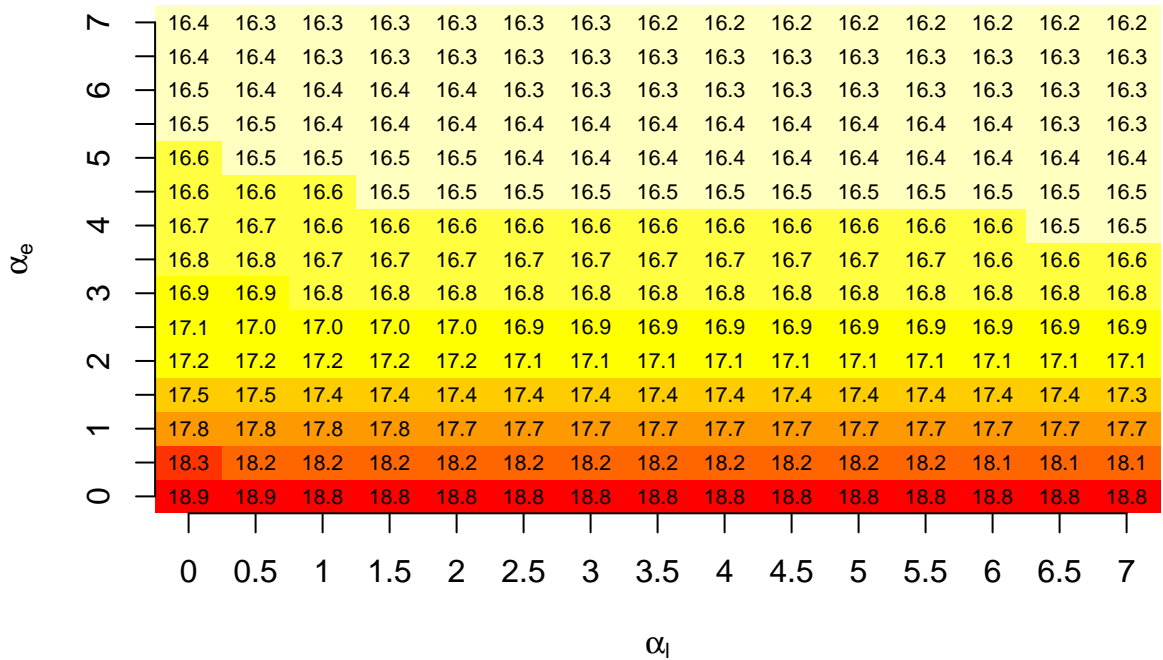

Code to create comparsion of different mean DAS trajectories over time with different values of $\alpha_e$. Can easily be modified to investigate different values of $\alpha_e$ and also vary $\alpha_l$:

```
time_plot <- seq(0, 7, by = 0.1)
unweighted <- calculate_mean_DAS(geemod=unweighted_gee)
weighted1 <- calculate_mean_DAS(geemod=weighted_gee)
weighted2<- calculate_mean_DAS(geemod=GEE_ANAR(alpha_e=4, alpha_l=0))
weighted3<- calculate_mean_DAS(geemod=GEE_ANAR(alpha_e=7, alpha_l=0))

plot(time_plot,unweighted,type="l", xlim=c(0,7),
     ylim=c(1,7),
     xlab="Time since diagnosis (years)",ylab="Mean Disease Activity Score", lwd=1)
lines(time_plot,weighted1,col=2, lwd=1)
lines(time_plot, weighted2, col="green", lwd=1)
lines(time_plot, weighted3, col="blue", lwd=1)

legend (2, 6,legend=c("Unweighted GEE", expression("IIW-GEE:" ~ paste(alpha[e], " = ", 0)),
                  expression("IIW-GEE:" ~ paste(alpha[e], " = ", 4)),
                  expression("IIW-GEE:" ~ paste(alpha[e], " = ", 7))),
       col=c("black", 'red', "green", "blue"),bty="n",lty=1)
```

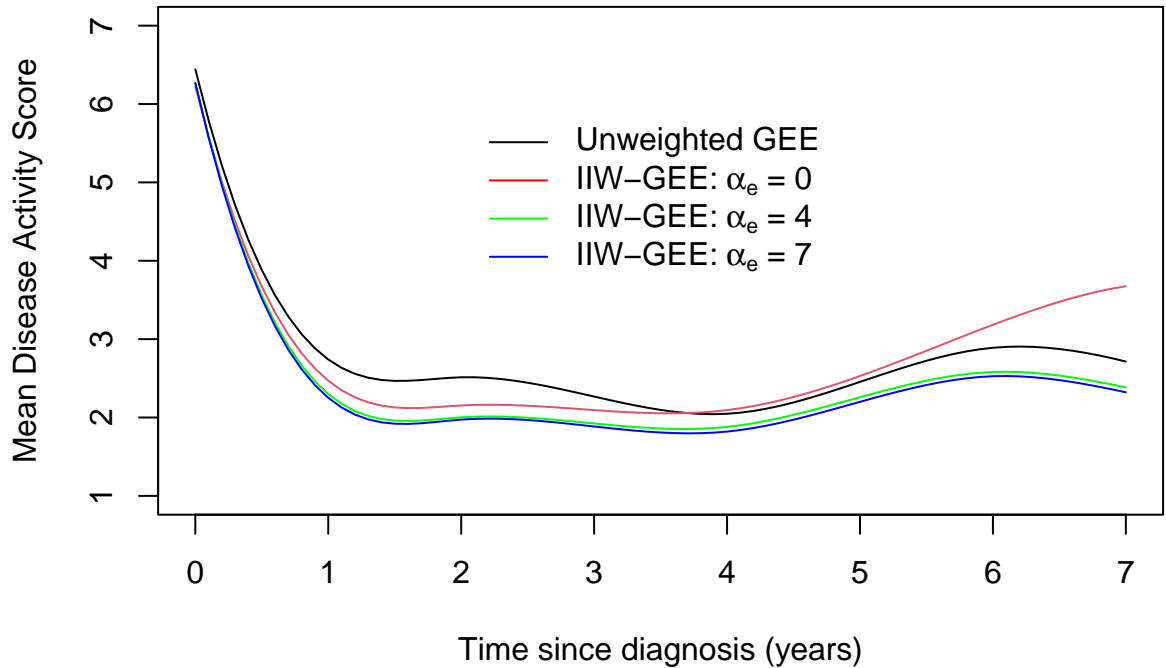